\documentclass[structabstract]{aa}  

\usepackage{txfonts}	
\usepackage{graphicx}

\usepackage{natbib}	

\begin{document}

\def\phcmskeV{\mbox{ph~cm$^{-2}$\,s$^{-1}$\,keV$^{-1}$}}
\def\phcmskeVsr{\mbox{ph~cm$^{-2}$\,s$^{-1}$\,keV$^{-1}$\,sr$^{-1}$}}
\def\keVcmssr{\mbox{keV\,cm$^{-2}$\,s$^{-1}$\,sr$^{-1}$}}
\def\phcms{\mbox{ph\,cm$^{-2}$\,s$^{-1}$}}
\def\ergcms{\mbox{erg\,cm$^{-2}$\,s$^{-1}$}}
\def\countcms{\mbox{count\,cm$^{-2}$\,s$^{-1}$}}
\def\lsim{\raisebox{-.5ex}{$\;\stackrel{<}{\sim}\;$}}
\def\gsim{\raisebox{-.5ex}{$\;\stackrel{>}{\sim}\;$}}
\newcommand{\mrm}[1]{\mathrm{#1}}
\newcommand{\dmrm}[1]{_{\mathrm{#1}}}
\newcommand{\umrm}[1]{^{\mathrm{#1}}}
\newcommand{\Frac}[2]{\left(\frac{#1}{#2}\right)}
\newcommand{\refeq}[1]{Eq.~(\ref{eq:#1})}
\newcommand{\refeqs}[2]{Eqs~(\ref{eq:#1}) and (\ref{eq:#2})}
\newcommand{\refeqss}[2]{Eqs~(\ref{eq:#1}) to (\ref{eq:#2})}
\newcommand{\reffig}[1]{Fig.~\ref{fig:#1}}
\newcommand{\reffigs}[2]{Figs.~\ref{fig:#1} and \ref{fig:#2}}
\newcommand{\reftab}[1]{Table~\ref{tab:#1}}
\newcommand{\reftabs}[2]{Tables~\ref{tab:#1} and \ref{tab:#2}}
\newcommand{\refsec}[1]{Sect.~\ref{sec:#1}}
\newcommand{\refsecs}[2]{Sects.~\ref{sec:#1} and \ref{sec:#2}}
\newcommand{\modif}[1]{#1}	

\title{\emph{INTEGRAL} hard X-ray spectra of the cosmic X-ray background and Galactic ridge emission}

\author{M. T\"urler\inst{1}\fnmsep\inst{2}
	\and M. Chernyakova\inst{3}
	\and T. J.-L. Courvoisier\inst{1}\fnmsep\inst{2}
	\and P. Lubi\'nski\inst{1}\fnmsep\inst{4}
	\and A. Neronov\inst{1}\fnmsep\inst{2}
	\and N. Produit\inst{1}\fnmsep\inst{2}
	\and R. Walter\inst{1}\fnmsep\inst{2}
}

\institute{ISDC Data Centre for Astrophysics, ch. d'Ecogia 16, 1290 Versoix, Switzerland\\
	\email{marc.turler@unige.ch}
	\and Geneva Observatory, University of Geneva, ch. des Maillettes 51, 1290 Sauverny, Switzerland
	\and Dublin Institute for Advanced Studies, 31 Fitzwilliam Place, Dublin 2, Ireland
	\and Nicolaus Copernicus Astronomical Center, Bartycka 18, 00--716 Warszawa, Poland
}

\date{Received 5 August 2009; accepted 4 January 2010}

\abstract
{
}{
We derive the spectra of the cosmic X-ray background (CXB) and of the Galactic
ridge X-ray emission (GRXE) in the $\sim$20--200\,keV range from the data of the
IBIS instrument aboard the \emph{INTEGRAL} satellite obtained during the four
dedicated Earth-occultation observations in early 2006.
}{
We analyze the modulation of the IBIS/ISGRI detector counts induced by the
passage of the Earth through the field of view of the instrument. Unlike
previous studies, we do not \modif{fix} the spectral shape of the various
contributions, but model instead their spatial distribution and derive for each
of them the expected modulation of the detector counts. The spectra of the
diffuse emission components are obtained by fitting the normalizations of the
model lightcurves to the observed modulation in different energy bins.
\modif{Because of degeneracy, we guide the fits with a realistic choice of the
input parameters and a constraint for spectral smoothness.}
}{
\modif{The obtained CXB spectrum is consistent with the historic \emph{HEAO-1}
results and falls slightly below the spectrum derived with \emph{Swift}/BAT. A
10\,\% higher normalization of the CXB cannot be completely excluded, but it
would imply an unrealistically high albedo of the Earth. The derived spectrum of
the GRXE confirms the presence of a minimum around 80\,keV with improved
statistics and yields an estimate of $\sim$0.6\,M$\dmrm{\odot}$ for the average
mass of white dwarfs in the Galaxy. The analysis also provides updated
normalizations for the spectra of the Earth's albedo and the cosmic-ray induced
atmospheric emission.}
}{
This study demonstrates the potential of \emph{INTEGRAL} Earth-occultation
observations to derive the hard X-ray spectra of three fundamental components:
the CXB, the GRXE and the Earth emission. Further observations would be
extremely valuable to confirm our results with improved statistics.
}


\titlerunning{\emph{INTEGRAL} hard X-ray spectra of the cosmic X-ray background and Galactic ridge emission}

\maketitle

\section{Introduction}
\label{sec:introduction}

\begin{figure*}[t]
\sidecaption		
\includegraphics[width=6cm]{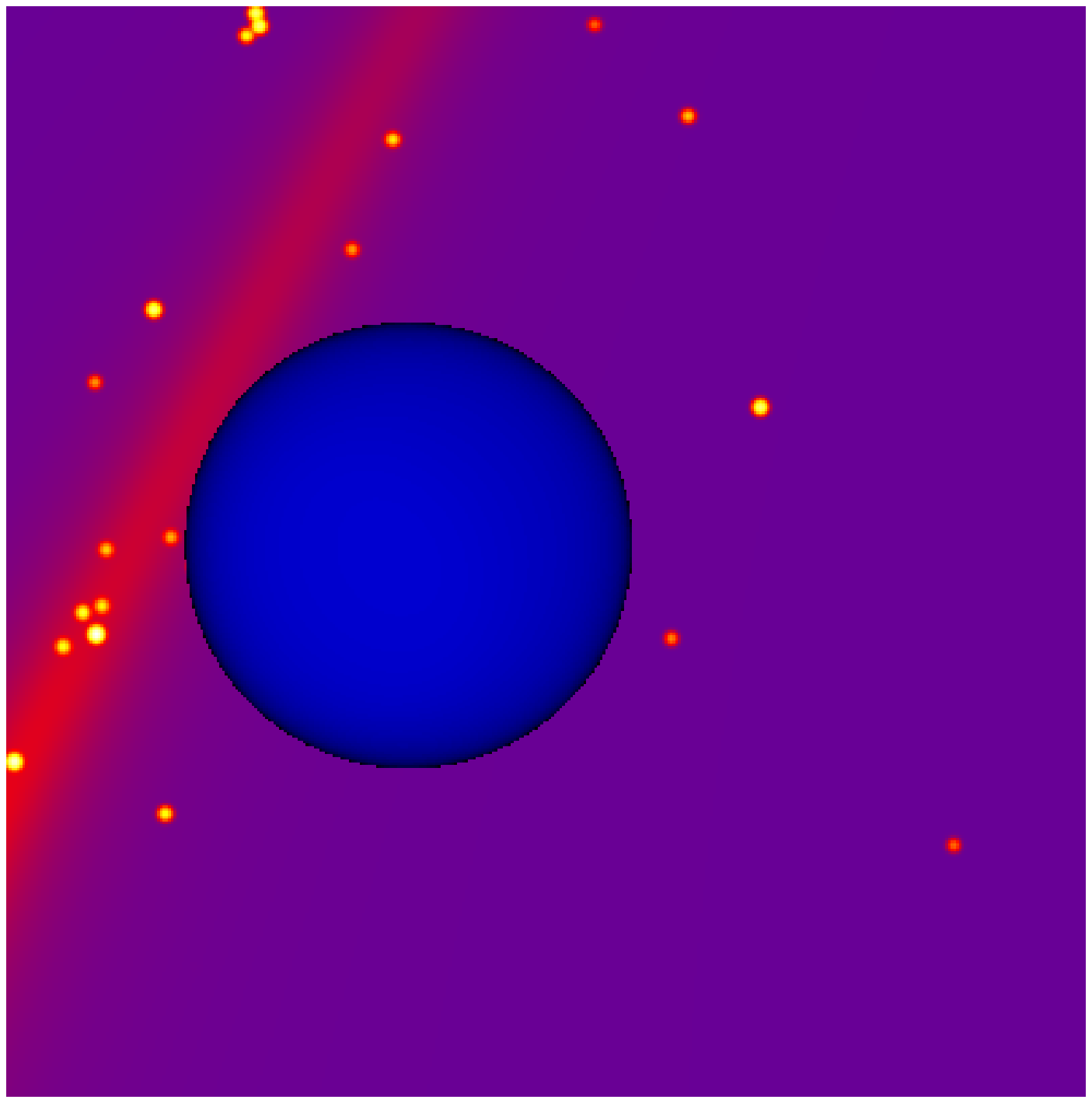}
\includegraphics[width=6cm]{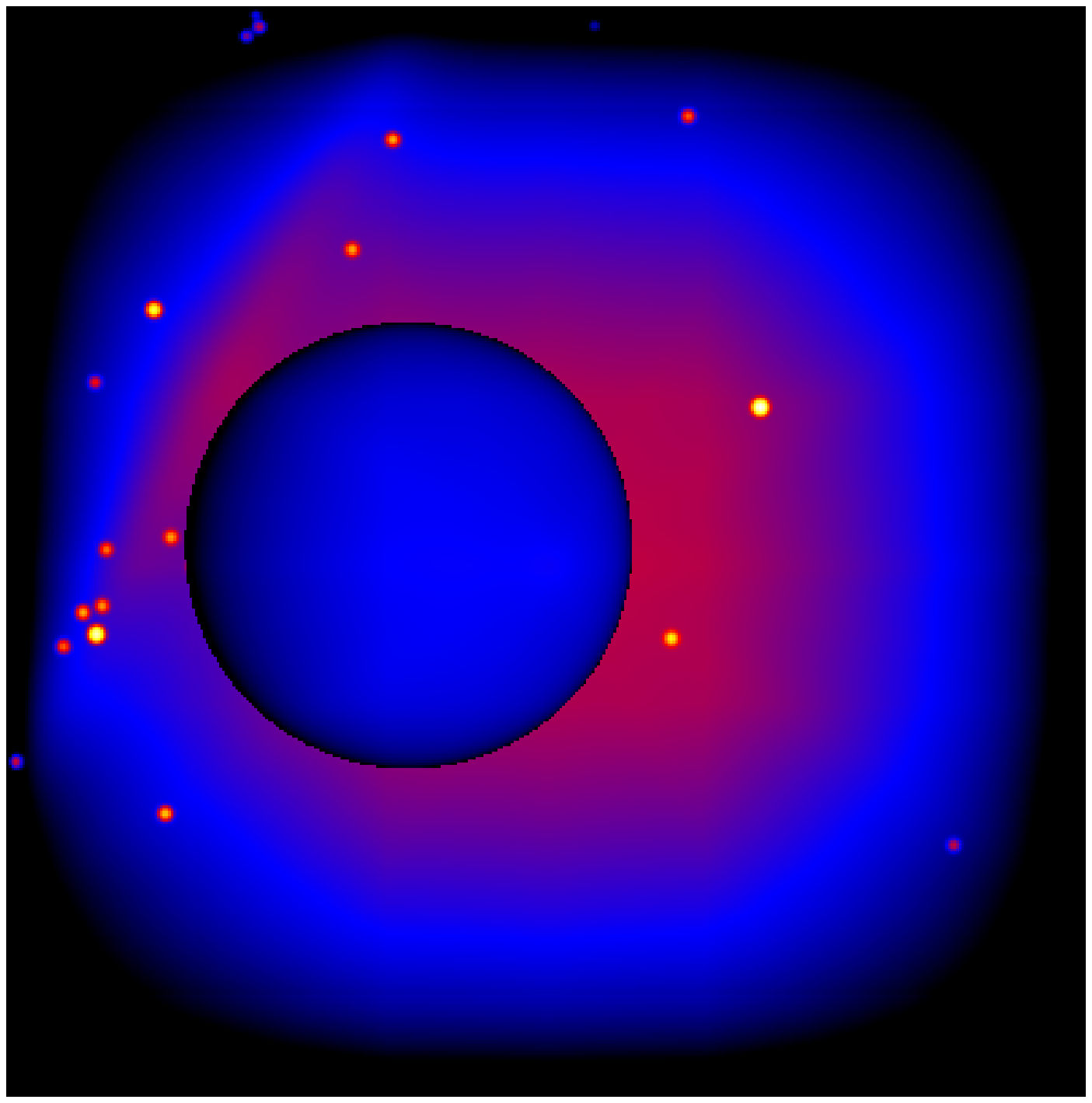}
\caption{\label{fig:ds9Image}
   Model IBIS/ISGRI images of the sky with (right) and without (left)
   instrumental vignetting effects (see \reffig{insEffects}). They show the
   geometry and relative intensity (on a logarithmic scale from dark and blue to
   red and bright areas) of the various components during the first EO at
   IJD\,=\,2216.04 as derived for channel 3 ($\sim$27\,keV). The bluish circle
   shows emission of the Earth occulting the diffuse sky background (purple),
   the Galactic ridge (red strip), and the point sources (bright dots). The
   instrumental background is ignored for clarity. The images extend over the
   partially coded FoV ($28.8\degr \times 29.2\degr$). \modif{Point sources} are
   convolved with a circular Gaussian typical for the instrumental resolution
   ($\sigma\!=\!0.1\degr$).
}
\end{figure*}

Although the cosmic X-ray background (CXB) was discovered before the cosmic
microwave background \citep{GGP62}, it is known in much less detail and its
spectral shape and normalization are still subjects of debate. This diffuse
emission is thought to be mainly due to unresolved active galactic nuclei (AGN)
extending to cosmological distances with a contribution from Type Ia supernovae
in the low-energy gamma-rays \citep{Z96}. Evidence for the AGN origin of the CXB
at energies below 10 keV comes from various X-ray mirror telescopes -- in
particular \emph{Chandra} and \emph{XMM-Newton} -- that were able to resolve up
to 80\,\% of the diffuse emission into point sources \citep[e.g.][]{BH05,GCH07}.
The amount of resolved sources decreases rapidly with energy though so that only
2.5\,\% of the diffuse background is resolved by the deepest survey yet in the
20--60\,keV range, at the peak of the CXB emission \citep{PWM08}. The
characterization of the actual spectral shape and normalization around this
emission bump is crucial to estimate the fraction of heavily absorbed
Compton-thick AGN thought to contribute significantly in this hard X-ray
spectral range \citep{UAO03,GCH07,SKR08,TUV09}.

The High Energy Astronomical Observatory 1 (\emph{HEAO-1}) is hitherto the only
satellite which had a dedicated mechanism to disentangle the CXB from the
instrumental background. By using a movable CsI crystal with a thickness of
5\,cm to cover part of the field of view (FoV), the \emph{HEAO-1} observations
of the mid-1970s are still the most accurate and reliable measurements of the
CXB spectral shape in the hard X-rays \citep{MBH80,KJG97,GMP99}. Without such a
masking mechanism in recent space missions, a practical way to study this
diffuse hard X-ray emission is to use the Earth as a screen occulting part of
the background sky. For pointing satellites in low orbits around the Earth, our
planet often crosses part of the field of view during normal operations. Such
events can be analyzed in detail to evaluate the CXB spectrum. This was done by
\citet{FOL07} for the \emph{BeppoSAX} mission and by \citet{AGS08} for the data
of the Burst Alert Telescope (BAT) aboard the \emph{Swift} spacecraft.

With its eccentric three-days orbit, the \emph{INTEGRAL} satellite \citep{WCD03}
is close to the Earth only during the perigee passage when the instruments are
not operating because of excessive background in the radiation belts. In order
to study the X-ray background, a series of four dedicated observations were
performed in January and February 2006. The Earth was allowed to pass through
the FoV of the instruments shortly after radiation-belt exit while the
spacecraft was aimed to point towards a fixed position in the sky. \citet{CSR07}
described these observations by all four instruments aboard \emph{INTEGRAL} in
great detail and studied how the passage of the Earth modulates the detector
counts by occulting part of the CXB.

The difficulty of using the Earth to shield the CXB comes from the fact that the
Earth is not dark in the hard X-rays. The emission from the Earth in the
20--200\,keV range consists of two major contributions: \modif{the reflection of
the CXB by the atmosphere and its Compton emission under the bombardment by
cosmic rays (CR)}. Disentangling the CXB occultation from the Earth emission is
challenging. \citet{CSR07} assumed the spectral shape of the CXB and of the two
Earth emission components and fitted their normalizations to the observed
amplitude of the Earth modulation in the data. The studies of the
\emph{BeppoSAX}/PDS measurements \citep{FOL07} and the \emph{Swift}/BAT 
observations \citep{AGS08} also relied on a priori assumptions on the spectral
shape of the CXB and the Earth emission.

We present here a completely different approach for the analysis of the same
\emph{INTEGRAL} observations as were used by \citet{CSR07}, but only focusing on
the data of the IBIS/ISGRI instrument \citep{ULD03}. Instead of fixing the
spectral shapes of the CXB and the Earth emission components, \modif{we aim to
derive them based on a detailed modeling of the spatial distribution of these
components and of all instrumental effects.} If the Earth surface brightness
differs significantly from the uniform CXB occultation, it is possible to
disentangle these components based on the recorded modulation when the Earth is
crossing the FoV. This has the potential to simultaneously derive the spectral
shape of the CXB and of the Earth emission from the observations.

Another difficulty of the analysis is the presence of the Galactic plane in the
border of the wide FoV of IBIS (see \reffig{ds9Image}). An empty extragalactic
field would have been ideal to study the CXB, but this was not possible due to
various scheduling constraints. This complication is however an opportunity to
study in addition the diffuse Galactic ridge X-ray emission (GRXE)
\citep[e.g.][]{RSG06,KRC07,BJR08}, which is also occulted by the passage of the
Earth.

The observational material and the analysis method are described in
\refsecs{data}{method}, respectively. We present the obtained spectra in
\refsec{results} and discuss them in comparison to previous results in
\refsec{discussion}, before concluding in \refsec{conclusion}. \modif{Unless
otherwise stated, the quoted errors are 1-$\sigma$ uncertainties, i.e. at the
68\,\% confidence level (CL).}

\section{Data}
\label{sec:data}
The data used here are the four Earth-occultation observations (EOs) conducted
by \emph{INTEGRAL} in January and February 2006 at the start of satellite
revolutions number 401, 404, 405 and 406. We refer the reader to the detailed
description of these observations in \citet{CSR07}. We focus our analysis on the
data of the IBIS/ISGRI gamma-ray imager that is best suited to study the
emission in the $\sim$20--200\,keV range.

Our work is based on the analysis of the modulation in full detector lightcurves
induced by the passage of the Earth through the FoV. These lightcurves are
obtained with the latest version of the \texttt{ii\_light} executable that will
be included in a forthcoming release of the Off-line Scientific Analysis (OSA)
software package provided by the \emph{INTEGRAL} Science Data Centre
\citep[ISDC,][]{CWB03}. They are corrected for instrumental dead time and the
effect of dead and noisy pixels, which amount to typically 5\,\% of all ISGRI
pixels. For each of the four similar observations, we extracted detector
lightcurves with a time binning of 300\,s in a series of 16 energy bins (see
\reftab{results}), carefully chosen to isolate instrumental emission features,
in particular the broad lines at 26 and 31\,keV from CdTe and the narrow lines
at 59\,keV from W, 75--77\,keV from Pb and 82--84\,keV from Bi \citep{TLB03}.

The detector lightcurves originally expressed in units of \countcms\ were
multiplied by 0.5\,($128\times 0.4$\,cm)$^2$ = 1310.72\,cm$^2$, which is the
area of the detector assumed by the standard ISGRI ancillary response file
(ARF), describing the energy dependence of the effective area of the instrument.
The factor 0.5 refers to the fraction of open coded mask elements, and 0.4\,cm
is the size of the $128\times 128$ ISGRI detector elements. Apart from this
change of unit to have full detector lightcurves, the only other manipulation of
the data was a cleaning of the lightcurves. This was done by removing points
with uncertainties of more than twice the average uncertainty and by iteratively
removing a few isolated outstanding points lying more than $3\,\sigma$ away from
the smoothed lightcurve with a smoothing window of 30 minutes (i.e. 6 time
bins).

In order to subtract from the lightcurves the contribution of point sources in
the FoV, we performed an image analysis separately for the four Earth
observations. This was possible since the drift of the satellite was moderate
despite the absence of star trackers during the pointings. The image analysis
was done in a standard way with the default background maps of OSA~7.0. We
searched for all sources previously detected by ISGRI including the new source
IGR~J17062--6143 already reported by \citet{CSR07}. We then selected all sources
that were detected with a significance of \modif{more than $2\,\sigma$} in the
22--60\,keV band. We chose this low significance threshold to minimize the
contribution of the known point sources to the GRXE and the CXB. Sources with
even less significance are more likely to be spurious and their global
contribution will mostly cancel out with fake negative sources. \modif{We tested
both a simple powerlaw and a bremsstrahlung model to fit the data. We found that
for most sources the bremsstrahlung model gives a better phenomenological
description of the data than a powerlaw because many sources have a convex
spectral shape in this energy range. The fluxes derived for the brightest
($>3\,\sigma$) sources in our sample are listed in \reftab{sources}.}

\begin{table}[tb]
\caption{\label{tab:sources}
   List of point sources detected in at least one of the four EOs.
}
\begin{flushleft}
\addtolength{\tabcolsep}{-1pt}
\begin{tabular}{@{}l@{~~~}cc@{~~~}r@{~~~}r@{~~~}r@{~~~}r@{}}
\hline
\hline
\rule[-0.5em]{0pt}{1.6em}
Source name& RA & Dec & EO\,1 & EO\,2 & EO\,3 & EO\,4 \\
           & \multicolumn{2}{c}{(~deg~)$^a$} & \multicolumn{4}{c}{(~$10^{-11}$\ergcms~)$^b$} \\
\hline
\rule{0pt}{1.2em}%
IGR J14471--6319    & 221.81 & $-$63.29 &   -- &   -- &   -- &  1.9\\
IGR J14515--5542    & 222.89 & $-$55.68 & 18.4 &   -- &   -- &   --\\
IGR J14532--6356    & 223.31 & $-$63.93 &   -- &   -- &   -- &  2.2\\
IGR J15094--6649    & 227.36 & $-$66.82 &  4.1 &   -- &   -- &  4.4\\
PSR B1509--58       & 228.48 & $-$59.14 &  7.2 &  7.5 &   -- &  5.6\\
IGR J15359--5750    & 233.97 & $-$57.83 &  2.9 &   -- &   -- &  1.4\\
H 1538--522         & 235.60 & $-$52.39 & 14.2 & 22.8 &   -- & 25.0\\
4U 1543--624        & 236.98 & $-$62.57 &   -- &  4.5 &  4.4 &   --\\
IGR J16167--4957    & 244.16 & $-$49.98 &   -- &  2.8 &   -- &  7.1\\
IGR J16207--5129    & 245.19 & $-$51.50 &  2.6 &  6.4 &   -- &   --\\
SWIFT J1626.6--5156 & 246.65 & $-$51.94 & 12.4 & 11.1 & 10.1 &  8.3\\ 
IGR J16283--4838    & 247.04 & $-$48.65 &  8.2 &   -- &   -- &   --\\
IGR J16318--4848    & 247.95 & $-$48.82 & 37.7 &   -- & 11.6 &  7.1\\
IGR J16320--4751    & 248.01 & $-$47.87 &  8.1 &   -- &   -- & 12.8\\
4U 1626--67         & 248.07 & $-$67.46 & 13.0 & 12.3 &  8.9 &  8.7\\
IGR J16377--6423    & 249.57 & $-$64.35 &   -- &  1.1 &   -- &   --\\
IGR J16393--4643    & 249.77 & $-$46.70 &   -- & 11.9 &   -- &  1.6\\
H 1636--536         & 250.23 & $-$53.75 &  7.1 &  2.3 &   -- &   --\\
IGR J17008--6425    & 255.20 & $-$64.43 &   -- &   -- &   -- &  0.9\\
XTE J1701--462      & 255.24 & $-$46.19 &   -- & 26.8 &   -- &   --\\
GX 339--4           & 255.71 & $-$48.79 &  9.5 & 28.7 & 40.9 & 44.3\\
IGR J17062--6143    & 256.57 & $-$61.71 &   -- &   -- &   -- &  4.5\\
NGC 6300            & 259.25 & $-$62.82 &  2.5 &  3.1 &  3.4 &  2.8\\
ESO 103--35         & 279.58 & $-$65.43 &   -- &   -- &  8.3 &   --\\
\hline
\end{tabular}\\
\rule{0pt}{1.0em}%
\textbf{Notes.}
$^{(a)}$ source catalog position in right ascension (RA) and declination (Dec).
$^{(b)}$ derived model fluxes in the 20--50\,keV band for each EO.
\end{flushleft}
\end{table}

\begin{table}[tb]
\caption{\label{tab:results}
   Numerical values of the obtained spectra shown in \reffig{plteeufs}.
}
\begin{flushleft}
\begin{tabular}{@{}rrrrrrrr@{}}
\hline
\hline
\rule[-0.5em]{0pt}{1.6em}
$E$~~~ & $\Delta E$ & $F\dmrm{sky}$ & $\Delta F\dmrm{sky}$ & $F\dmrm{ear}$ & $\Delta F\dmrm{ear}$ & $F\dmrm{gal}$$^c$ & $\Delta F\dmrm{gal}$$^c$ \\
\multicolumn{2}{c}{(~keV~)$^a$} & \multicolumn{6}{|c}{(~\keVcmssr~)$^b$} \\
\hline
\rule{0pt}{1.2em}%
 21.14 &  1.44 & 39.57 &  2.53 &  7.05 &  1.58 & 18.09 & 2.07 \\ 
 24.01 &  1.44 & 42.79 &  1.92 &  8.63 &  1.47 & 19.94 & 0.55 \\ 
 27.36 &  1.91 & 43.84 &  1.81 & 11.23 &  1.31 & 19.26 & 4.88 \\ 
 31.19 &  1.91 & 42.90 &  1.44 & 13.79 &  1.21 & 16.62 & 2.75 \\ 
 35.02 &  1.91 & 41.08 &  2.45 & 17.02 &  1.60 & 14.35 & 4.26 \\ 
 38.85 &  1.91 & 39.08 &  2.55 & 20.31 &  1.88 & 12.12 & 2.63 \\ 
 43.16 &  2.39 & 35.42 &  2.42 & 23.87 &  2.46 &  9.80 & 0.84 \\ 
 47.95 &  2.39 & 38.67 &  1.95 & 26.82 &  2.16 &  9.70 & 1.35 \\ 
 52.73 &  2.39 & 36.24 &  2.42 & 29.30 &  2.00 &  8.33 & 1.40 \\ 
 58.00 &  2.87 & 30.88 &  3.41 & 30.85 &  2.14 &  7.72 & 1.37 \\ 
 63.74 &  2.87 & 28.86 &  2.81 & 31.18 &  2.48 &  7.17 & 1.78 \\ 
 71.88 &  5.27 & 28.92 &  3.34 & 32.47 &  3.11 &  6.46 & 1.31 \\ 
 83.37 &  6.22 & 30.01 &  6.35 & 33.37 &  3.23 &  6.17 & 0.80 \\ 
 94.86 &  5.27 & 27.12 &  4.89 & 34.01 &  2.71 &  6.64 & 0.84 \\ 
112.57 & 12.45 & 24.82 &  5.94 & 35.04 &  3.90 &  6.68 & 0.99 \\ 
162.35 & 37.34 & 30.65 &  9.55 & 37.85 &  7.46 &  7.93 & 2.86 \\ 
\hline
\end{tabular}\\
\rule{0pt}{1.0em}%
\textbf{Notes.}
$^{(a)}$ Central energy, $E$, and half-width, $\Delta E$, of the bins.
$^{(b)}$ Fluxes, $F$, and statistical uncertainties, $\Delta F$, for the sky
background (sky), the Earth (ear) and the Galaxy (gal).
$^{(c)}$ On average over the region $320\degr<l<340\degr$ and $|b|<5\degr$.
\end{flushleft}
\end{table}

\section{Method}
\label{sec:method}
The detector lightcurves described above were modulated by the passage of the
Earth through the FoV of IBIS. Our approach was to model these observations in
detail to derive the expected modulation of the detector counts for each
emission component on the sky. This resulted in a series of model lightcurves in
different energy bins for each emission component and each of the four EOs. We
then fitted the normalizations of these model lightcurves to the observed
detector counts to derive the actual contribution of the diffuse emission
components.

This method requires the knowledge of the spacecraft position and attitude with
respect to the Earth and to the background sky at any time, a description of the
spatial distribution on the sky of the various emission components and also an
accurate description of the IBIS/ISGRI instrumental characteristics. These
aspects are described in the three subsections below. The generation of the
model lightcurves is described in \refsec{lightcurves}, whereas the actual
spectral fitting procedure is the subject of \refsec{spectral}.

\subsection{Satellite position and attitude}
\label{sec:satellite}

To construct the images of the sky corresponding to each of the four Earth
observations as illustrated in \reffig{ds9Image}, we needed to know the exact
attitude of the satellite and its distance to the Earth at any time. This
information can be extracted from auxiliary files provided by the mission
operation centre (MOC) in Darmstadt. It was used to compute the position of the
Earth center, the position of the geographic and magnetic poles, and the
apparent radius of the Earth \modif{as a function of time, all expressed in
degrees, using} the IBIS/ISGRI instrument coordinates $(Y,Z)$. \modif{The
spacecraft was close enough to the Earth at the beginning of the observation for
the planet's sphericity to slightly affect its apparent radius. This was
properly taken into account, as well as the $\sim$100\,km of obscuring
atmosphere in the hard X-rays mentioned by \citet{CSR07}. The magnetic pole in
the Northern hemisphere is set to its 2005 position of 82.7\degr\,N,
114.4\degr\,W.}

\subsection{Spatial distribution of components}
\label{sec:spatial}
\modif{Although we were only interested in the temporal modulation of counts on
the full detector area, we needed a sufficiently precise} description of the
spatial distribution of the emission components. We chose to \modif{define all
of} them by analytical functions that are described below and are illustrated in
\reffig{ds9Image}.

The simplest component is the CXB that we assumed to be uniform on the sky.
Although there is evidence that the CXB has some large-scale intensity
variations \citep[e.g.][]{BCK02,RMS08}, they are small in amplitude
(\lsim\,2\,\%) and it would be very difficult to evaluate and account for a
possible non-uniformity so close to the Galactic bulge.

The GRXE was modeled with two perpendicular Lorentzian functions aligned with
the Galactic coordinates. The full-width at half maximum (FWHM) of the
Lorentzians are of $21\degr$ \citep[Fig.~7]{KRC07} and $1.2\degr$
\citep[Fig.~5]{RSG06} respectively along the Galactic longitude, $l$, and
latitude, $b$. The Lorentzian's maximum are at the Galactic center with a slight
latitude displacement of $b=-0.15\degr$ as measured by \citet{RSG06}. \modif{As
shown by \citet{KRC07}, this distribution matches well the \emph{COBE}/DIRBE map
at 4.9\,$\mu$m, which was used by \citet{BJR08} as a template for the GRXE below
120\,keV.}

\modif{We took special care to define the spatial distribution of the Earth's
emission. There are two distinct components to be taken into account: the CXB
reflection by the Earth \citep{CSS08} and the CR-induced atmospheric emission
\citep{SCS07}. The authors of these studies performed Monte-Carlo simulations to
derive both the spectrum and the surface brightness of the Earth emission. We
used the latter results as a precise determination of the expected image of the
Earth at hard X-rays.}

\modif{\citet{CSS08} found that the X-ray albedo of the Earth is limb-darkened
at lower energies and limb-brightened at higher energies. At energies below
$\sim$100\,keV -- where this component dominates the Earth emission -- the
emission is found to be limb-darkened, but slightly less than for a sphere
emitting black-body radiation. Such an object would have a linear dependence of
the flux with $\mu\!\equiv\!\cos{\theta}$, where $\theta$ is the zenith angle,
i.e. the angle between the line-of-sight and the normal to the surface. For the
Earth albedo below $\sim$100\,keV they instead found an angular dependence of
the reflected flux that can be approximated by $F(\mu)\propto
\mu\,(1-0.5\,\mu)$. We used this equation to define the Earth albedo component.}

\modif{We modeled the CR-induced emission of the Earth's atmosphere
according to \citet[Eq.~(7)]{SCS07}. By setting the solar modulation potential
to $\phi\!=\!0.5$ -- corresponding to the solar minimum during the EOs of 2006
-- we can simplify this equation as:}
\begin{equation}
\label{eq:cr_emis}
C\propto \mu\,(1+\mu)\,\left(\,1+(R\dmrm{cut}/3.2)^2\,\right)^{-0.5},
\end{equation}
\modif{where $\mu$ is as defined above and $R\dmrm{cut}$ is the geomagnetic
cut-off rigidity. In the dipole approximation of the Earth's magnetic field,
the latter depends mainly on the geomagnetic latitude $\lambda\dmrm{m}$ as
$R\dmrm{cut}\simeq 14.5\,\cos^4{\lambda\dmrm{m}}$\,GV \citep{SS05}. The
resulting atmospheric emission of the Earth is a combination of relatively
strong limb-darkening from the $\mu$-dependence in \refeq{cr_emis} with enhanced
emission at the magnetic poles from the $\lambda\dmrm{m}$-dependence.}

\modif{We note that we took into account for both Earth emission components the
distortion of the surface brightness related to the fact that only a portion of
the Earth's hemisphere can be seen when the spacecraft is relatively close to
the planet.}

\begin{figure}[t]
\includegraphics[bb=16 150 430 700,clip,width=\hsize]{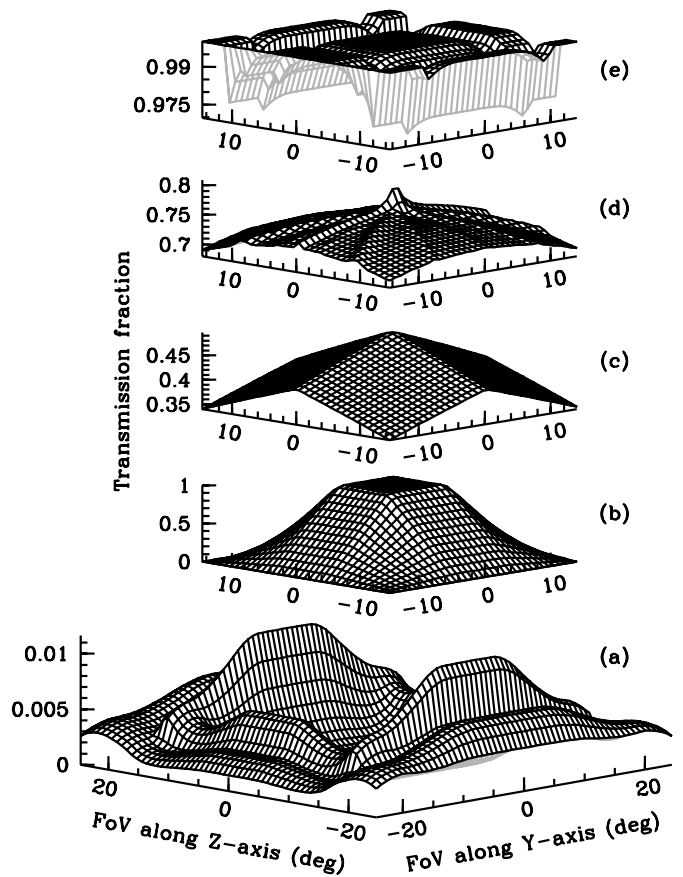}
\caption{\label{fig:insEffects}
   Surfaces representing the five IBIS/ISGRI vignetting effects that affect the
   incoming radiation until it reaches the detector plane. The effects are those
   corresponding to channel 12 ($\sim$72\,keV) and are:
   (a) the IBIS tube transparency to off-axis radiation,
   (b) the energy independent exposure map,
   (c) the effective coded-mask transparency,
   (d) the transmission of the Nomex structure supporting the mask, and
   (e) the absorption of the ISGRI spider beams.
   The total vignetting effect is the product of the effects (b) to
   (e) with the addition of effect (a), which extends well outside
   the partially coded FoV.
}
\end{figure}

\subsection{Instrumental characteristics}
\label{sec:instrumental}

As we wanted to fit real detector lightcurves with model lightcurves we needed
to take into account the instrumental characteristics of the telescope in the
\modif{modeling}. This does not include detector responses, but all effects
attenuating the incoming photon field on its path from outside the telescope
until reaching the detector plane. We identified five effects that affected the
detector illumination depending on the direction of the incoming radiation and
sometimes on its energy. The most obvious effect is the attenuation due to the
coded-mask elements which block out about half of the incoming radiation. The
second effect is the non-uniform exposure map which is caused by a partial
illumination of the detector for a source outside of the fully coded FoV.
Another vignetting effect is induced by the Nomex honeycomb structure that
supports the coded mask. The two last effects are due to the IBIS/ISGRI spider 
beams separating the eight detector modules and to the lead shielding of the
IBIS telescope tube. The aluminum spider results in opacity at the lowest
energies, and the lead shielding of the tube becomes transparent at the highest
energies.

The five effects mentioned above are included in the standard IBIS/ISGRI
software for image reconstruction and spectral extraction, but as we
worked directly with the detector lightcurves, we needed to account for these
effects independently. Their \modif{modeling} as used in this work is described
below and is illustrated in \reffig{insEffects}, while the overall vignetting
effect is shown in \reffig{ds9Image}.

The coded-mask transparency is ideally of $0.5$ since there are as many elements
open as closed. This is, however, only true at the center of the FoV. For
off-axis sources there is an additional attenuation due to the thickness of
16\,mm of the mask elements, which project a wider shadow on the detector for
increasing off-axis angles. As the elements are made of tungsten -- a strongly
absorbing material -- it is fair to assume the elements to be completely opaque
in the energy range considered here. As we were only interested in the net
effect over the full detector plane and in a simple analytical description we
approximated the mask pattern as a giant chessboard of 46 equally-sized square
elements on a side of 1064\,mm. We then properly computed the additional shadow
from radiation that crossed the border of the mask elements and did not fall
onto the shadow of other elements.

The exposure map of IBIS/ISGRI is basically very simple with a value of 1 in the
fully coded FoV and a linear decrease to zero in the partially coded FoV, except
in the corners of the image where the decrease is quadratic. \modif{When
we modeled this, we properly took into account the disposition of the eight
modules of the IBIS/ISGRI detector and the two-pixel wide space between them.}

The Nomex structure supporting the coded mask of IBIS is absorbing part of the
incoming photons. This is corrected for in the OSA software by off-axis
efficiency maps depending on energy. \modif{In 2006, at the time of the EOs,
these maps were still an approximation with only a dependence on the off-axis
angle. We used the new maps introduced in the OSA 6.0 release that do include an
additional azimuthal dependence due to the alignment of the walls of the
hexagonal tubes that the honeycomb structure is made of and also a correction
for the tubes pointing $\sim0.5\degr$ away from the center of the FoV; a
misalignment likely due to on-ground manipulations of the spacecraft.} We note
that the attenuation by the cosine of the off-axis angle is included in these
maps.

The eight IBIS/ISGRI detector modules are separated by an aluminum structure
called the ISGRI spider. It is made of one beam along the $Z$ axis and three
perpendicular beams. According to the IBIS experiment interface document part B
(EID-B) the beams have a trapezoidal section with a height of $48$\,mm a base of
$9.5$\,mm and a wall angle of $4\degr$ resulting in an upper width of $2.8$\,mm.
The wall angle ensures that the spider is not casting a shadow in the
fully coded FoV. However, in the partially coded FoV, the spider can mask up to
two rows of ISGRI pixels. We modeled this effect in detail for each energy bin
based on the corresponding attenuation length of Al. We found that for some
specific off-axis directions, the ISGRI spider can result in an attenuation of
the radiation on the detector plane of up to $\sim 5$\,\% in the lower energy
bins.

The last instrumental effect we considered is the IBIS telescope tube
transparency. At higher energies, the tube becomes transparent to radiation from
outside the fully coded FoV, giving an additional contribution to the detector
lightcurves. The tube is made of two vertical walls perpendicular to the $Y$
axis and of two walls inclined with an angle of $3.47\degr$ transverse to the
$Z$ axis. The walls are shielded with glued lead foils. The thickness of the Pb
sheets for each wall is reported in Table~3.2.8.1 of the EID-B (issue 7.0). We
used the values for the four upper sheets of the wall that are relevant for the
calculation of the tube transparency for an off-axis angle up to about 45\degr.
The calculation was done carefully, avoiding radiation further blocked by other
parts of the ISGRI collimating system: the 1\,mm thick W shield of the ISGRI
hopper and the 1.2\,mm thick W-strips of the side shielding of the mask. The
tube transparency is neglectable at the lower energies, but reaches $\sim 1\,\%$
in channel 12 (see \reffig{insEffects}) near the Pb attenuation length edge at
about 75--80\,keV and $\sim 6\,\%$ in the last energy bin. Although this might
seem unrelevant, the cumulated effect on the detector plane from a wide region
outside the FoV is actually far from negligible at these energies.

\begin{figure}[t]
\includegraphics[width=\hsize]{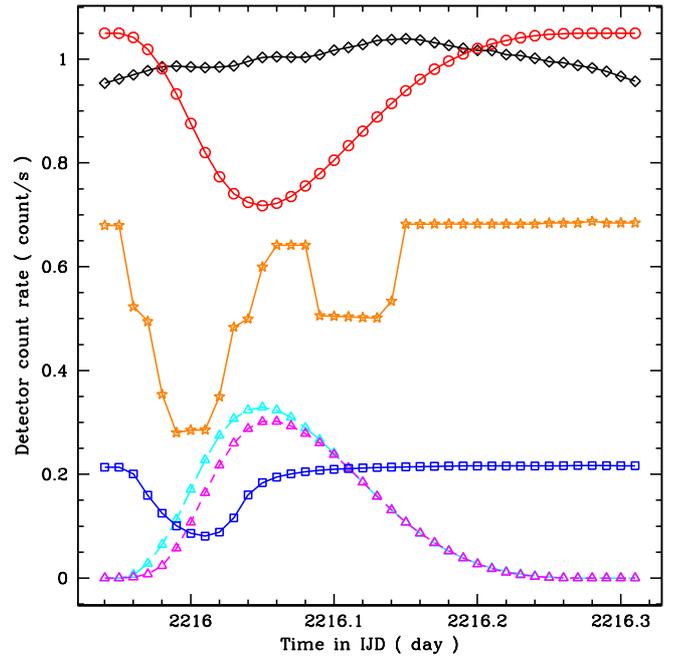}
\caption{\label{fig:pltLC}
   Model lightcurves of each component for channel 3 ($\sim$\,27\,keV) of the
   first EO. The detector count rate values are the effective contribution of
   the point sources (orange stars). The other components are the sky background
   (red circles), the GRXE (blue squares), and the Earth \modif{CXB reflection}
   (cyan triangles, long-dashed) and the \modif{CR-induced emission} (magenta
   triangles, short-dashed). For these diffuse components the detector count
   rate corresponds to an incident radiation of 40\,count\,s$^{-1}$\,sr$^{-1}$.
   For the instrumental background (black diamonds) we show the relative
   modulation, $M\dmrm{ins}(t)/\overline{M\dmrm{ins}}$, derived from the SPI/ACS
   corresponding to a detector count rate of 10\,count\,s$^{-1}$.
}
\end{figure}

\subsection{Construction of model lightcurves}
\label{sec:lightcurves}
The next step was to construct model lightcurves describing the modulation of
the radiation of each component described in \refsec{spatial} as induced by the
passage of the Earth through the IBIS/ISGRI FoV. A complete set of model
lightcurves is shown in \reffig{pltLC}.

\subsubsection{Extended components}
\label{sec:extended}

For the diffuse components -- the CXB, the GRXE and the two different Earth
emission components -- we constructed the model lightcurves by generating a
series of images of the sky at different times based on the attitude of the
spacecraft and the position of the Earth in the FoV (see \refsec{satellite}).
For each individual component, we considered only its own contribution and the
instrumental vignetting effects (see \refsec{instrumental}) attenuating the
count rates on the detector plane. The sum of the pixels in the images generated
for different times during the Earth occultation defines the model lightcurve
for a given component. Because the instrumental attenuation is energy dependent,
we constructed these model lightcurves for each energy bin and for each of the
four EOs because of the slightly different pointing directions with respect to
the Galactic ridge and Earth positions. As high-energy radiation from outside
the field of view also contributes to the detector counts (see
\refsec{instrumental}), the simulated images were defined on a wide area
extending 30\degr outside of the actual FoV of IBIS/ISGRI ($|Y|<44.4\degr$ and
$|Z|<44.6\degr$).

The normalization of the diffuse components in the input images -- without
vignetting effects -- was set to 10\,count\,s$^{-1}$\,sr$^{-1}$. For the Earth
emission components, this is the average intensity over the Earth disk, while it
is the average in the area defined by $320\degr<l<340\degr$ and $|b|<5\degr$ for
the GRXE. After attenuation by the instrumental effects described in
\refsec{instrumental} the actual detector count rate was typically reduced by an
order of magnitude (see \reffig{pltLC}).

\subsubsection{Point sources}
\label{sec:src}
In addition to the lightcurves constructed for the extended components we also
generated one lightcurve in each energy bin for the point sources in the FoV.
The time modulation is step-like in this case due to the abrupt disappearance of
a source when it gets occultated by the Earth. The count rate used for each
\modif{source detected with a significance of more than $2\,\sigma$ was derived
from a bremsstrahlung} fit to its observed IBIS/ISGRI spectrum (see
\refsec{data}) divided by the fraction of time during which the source was not
occulted. These counts were then assigned to the corresponding source position
in the simulated sky images, and detector counts were obtained by summing-up the
image pixels after application of the instrumental vignetting effects. We did
this at different times during the passage of the Earth through the FoV to get
model detector lightcurves. As the set of model lightcurves for point sources at
different energies was based on the actual data collected during each EO, they
were considered as a fix contribution to the detector lightcurves.

\subsubsection{Instrumental background}
\label{sec:ins}
The last but the dominant contributor to the observed detector lightcurves is
the instrumental background. The time variability of this component depends on
the cosmic particle environment \modif{and the induced radioactive decay of the
spacecraft's material}. The particle environment is well monitored by the
anti-coincidence shield (ACS) of the spectrometer SPI, the other gamma-ray
instrument of \emph{INTEGRAL} \citep{VRS03}. We found \modif{good} evidence that
the IBIS/ISGRI detector lightcurves are indeed \modif{following} the variations
recorded by the SPI ACS. To estimate the actual relationship between the count
rates in the ACS and in the IBIS/ISGRI detector in each of the considered
spectral bins, we used the extragalactic observations of revolution 342 (Her X-1
and XMM LSS). These observations, away from bright hard X-ray sources, were
taken about six months before the EOs and have the particularity of including a
solar flare at the start of the revolution, resulting in important correlated
variations in the SPI ACS and the ISGRI detector counts during the 12-hour decay
of the flare between \emph{INTEGRAL} Julian dates (IJD = JD $-\ 2\,451\,544.5$)
of 2040.0 and 2040.5. This relationship is characterized by the slope $\alpha$
of a linear fit of the ISGRI counts versus the SPI/ACS counts. This slope is
likely to change from one observation to the other because the orientation of
the spacecraft with respect to the solar radiation and particle flux will change
the effective areas of both the SPI/ACS and the IBIS/ISGRI detectors in a
complex manner. However, the energy dependence of the slope $\alpha(E)$ for
different ISGRI energy bins is expected to be rather stable. \modif{We used this
energy dependence of $\alpha$ as an indication of the amount of SPI/ACS
modulation expected in the ISGRI detector lightcurves of the EOs. The model
lightcurve for the variations of the instrumental background was thus
constructed based on those of the SPI/ACS as:}
\begin{equation}
\label{eq:ins}
M\dmrm{ins}(E,t) = \alpha(E)\,\left( C\dmrm{ACS}(t)-\overline{ C\dmrm{ACS}} \right)\,,
\end{equation}
where $C\dmrm{ACS}(t)$ is the SPI/ACS count rate lightcurve measured during the
EOs.

\begin{figure}[t]
\includegraphics[width=\hsize]{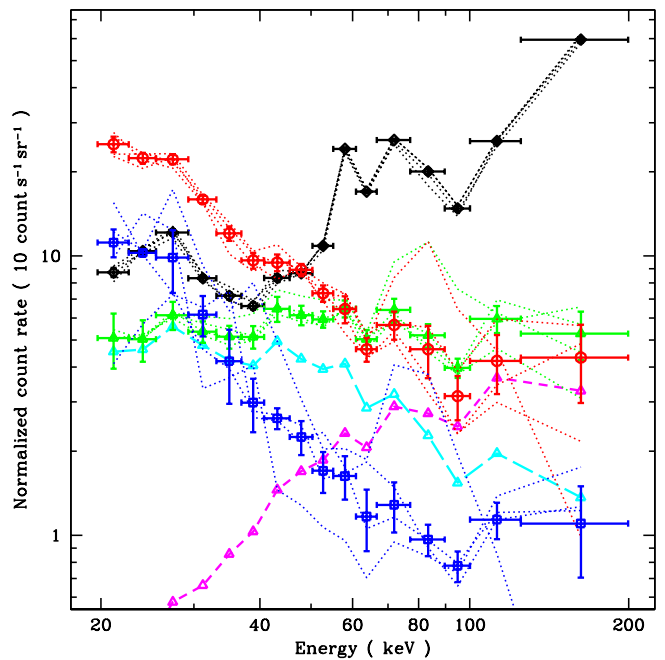}
\caption{\label{fig:pltSpectra}
   IBIS/ISGRI count rate spectra of each model component derived from the
   detector lightcurves of the \emph{INTEGRAL} EOs. The values are
   vignetting-corrected count rates per channel in units of
   10\,count\,s$^{-1}$\,sr$^{-1}$, except for the instrumental background (black
   diamonds) for which they are actual detector count rates in count\,s$^{-1}$.
   For the CXB (red circles), the GRXE (blue squares) and the total Earth
   emission (green triangles), the dotted lines of the same color show the
   average spectra obtained for the four independent EOs. The relative
   contributions to the Earth emission from the \modif{CXB reflection} (cyan
   triangles, long-dashed) \modif{and the CR scattering in the atmosphere}
   (magenta triangles, short-dashed) are also shown.
}
\end{figure}

\begin{figure*}[t]
\includegraphics[width=0.33\hsize]{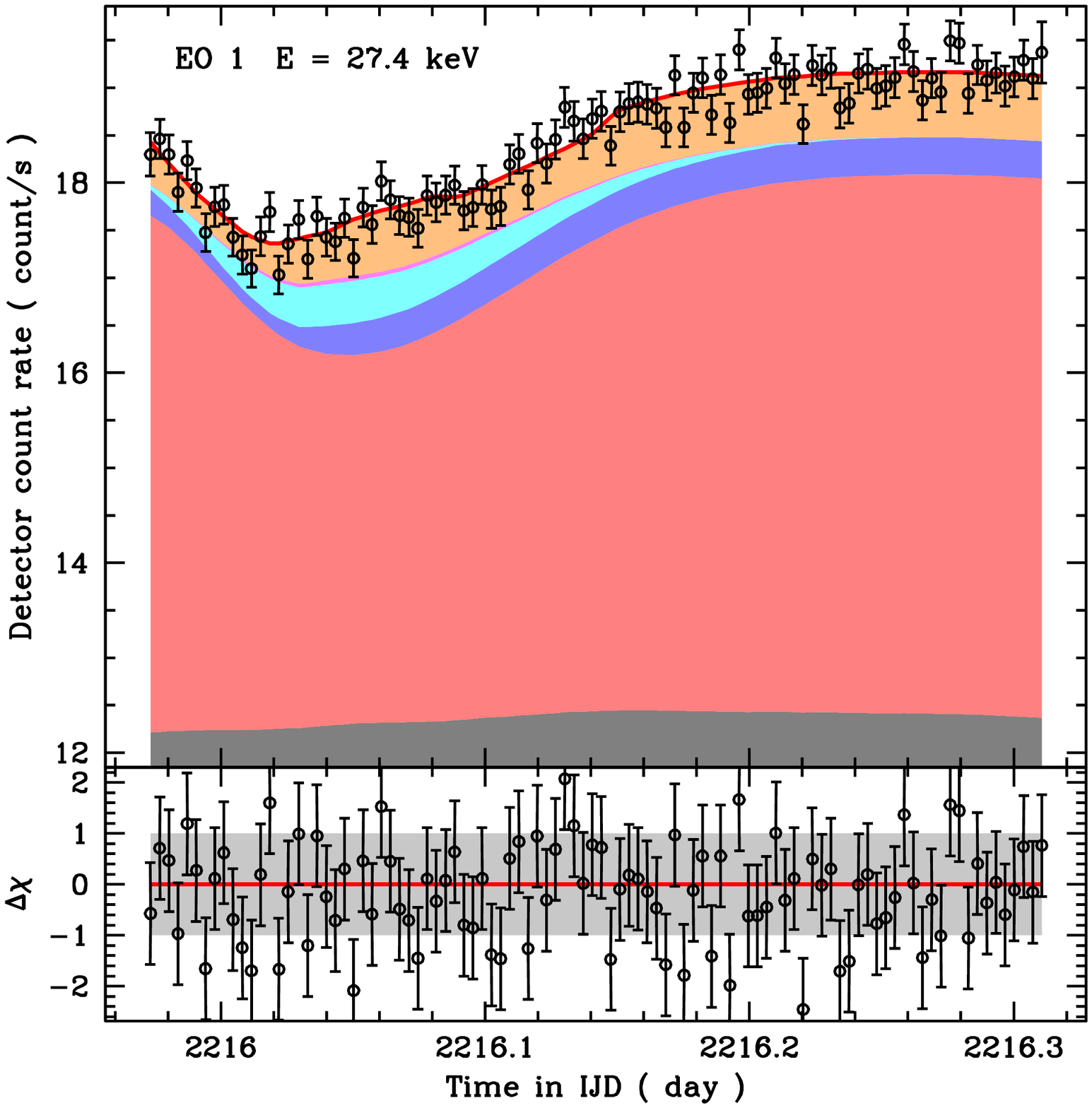}
\includegraphics[width=0.33\hsize]{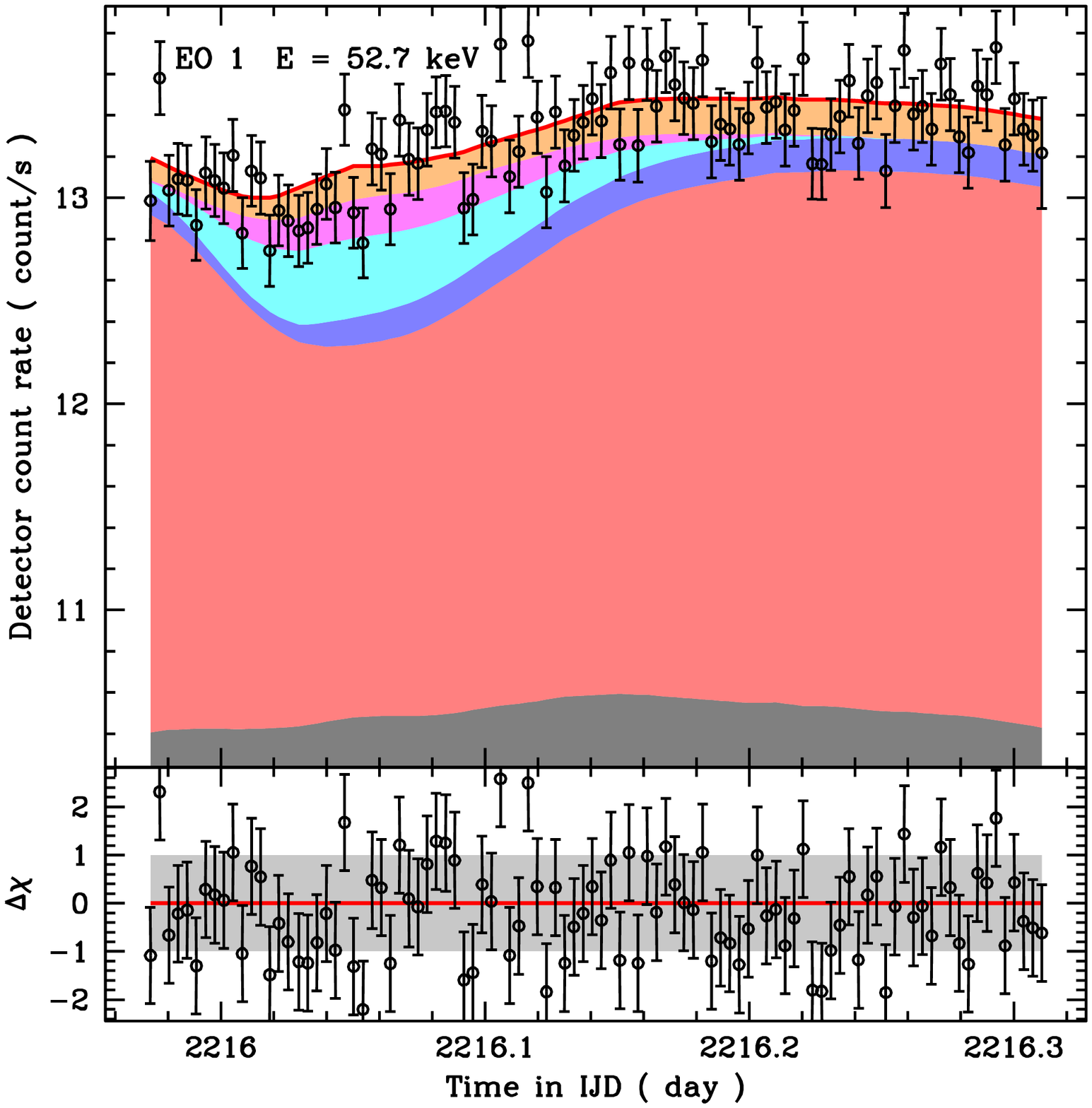}
\includegraphics[width=0.33\hsize]{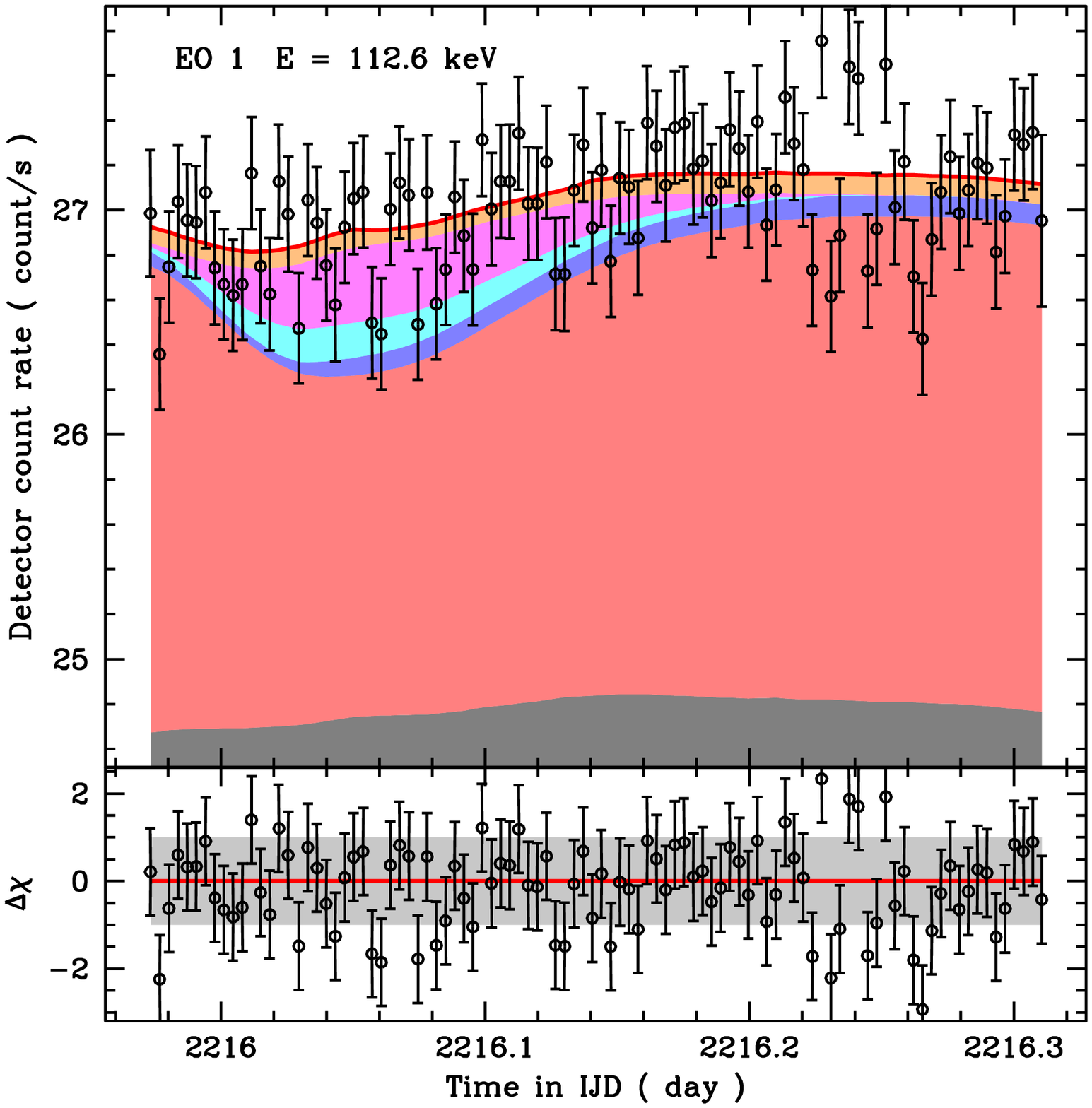}
\caption{\label{fig:pltFit}
   Examples of detector lightcurve fits (\emph{upper panels}) and associated residuals
   (\emph{lower panels}) for the first EO at three representative energies:
   $\sim$27\,keV (channel 3), $\sim$52\,keV (channel 9) and $\sim$112\,keV
   (channel 15), \emph{from left to right}. The reduced $\chi^2$ value of the best-fit
   curve (red line) is \modif{0.97, 1.08 and 1.06} respectively. The
   contribution of different components is shown by different colors with the
   same coding as in \reffigs{pltLC}{pltSpectra}. From bottom to top we add to
   the instrumental background (grey) -- modulated by the SPI/ACS lightcurve and
   a possible trend with time -- the sky background (red), the GRXE (blue), the
   \modif{albedo} (cyan) and \modif{atmospheric} (magenta) Earth emissions, and
   the fixed contribution from point sources (orange).
}
\end{figure*}

\subsection{Spectral fitting}
\label{sec:spectral}

We described above the construction of the model lightcurves $M_i(t)$ shown in
\reffig{pltLC} for each component $i$, which are the \modif{SPI/ACS-related}
variations of the instrumental background (ins), the sky background (sky), the
GRXE (gal), \modif{the Earth's albedo (alb), the atmospheric CR-induced emission
(atm)}, and the point sources (src) in the FoV. The next step is to adjust these
model lightcurves to the observed detector lightcurve $D(t)$ in a given energy
band with a least-square fit. This is done by the following linear relation:
\begin{eqnarray}
\label{eq:fit}
D(t) \approx & a\dmrm{ins} + b\dmrm{ins}\,\frac{t\,-\,\overline{t}}{t\dmrm{end}\,-\,\overline{t}} + c\dmrm{ins}\,M\dmrm{ins}(t) + c\dmrm{sky}\,M\dmrm{sky}(t) + \nonumber\\
             & c\dmrm{gal}\,M\dmrm{gal}(t) + c\dmrm{alb}\,M\dmrm{alb}(t) + c\dmrm{atm}\,M\dmrm{atm}(t) + M\dmrm{src}(t) \,,
\end{eqnarray}
\modif{where $a\dmrm{ins}$, $b\dmrm{ins}$ and the five $c_i$ are the seven free
parameters of the fit, scaling the model lightcurves $M_i(t)$ to best match the
observations $D(t)$. The $a\dmrm{ins}$ parameter describes the average value of
the instrumental background, whereas the $b\dmrm{ins}$ and $c\dmrm{ins}$
parameters model its variations. The parameter $b\dmrm{ins}$ allows us to
account for a linearly increasing or decreasing trend of the instrumental
background during the observation, centered on a time $\overline{t}$ and ending
at a time $t\dmrm{end}$. This turned out to have an important effect at energies
above 60\,keV (see \refsec{degeneracy}). Short-term variations of the
instrumental background $M\dmrm{ins}(t)$ are derived from the simultaneous
SPI/ACS lightcurve according to \refeq{ins} and are scaled by the parameter
$c\dmrm{ins}$. The four other $c_i$ parameters} are the count rates of the
various diffuse emission components on the sky, already corrected for
instrumental vignetting effects. Finally, the model lightcurve for the point
sources in the FoV, $M\dmrm{src}(t)$, was not scaled as it already represents
effective count rates in the detector.

By fitting the observed lightcurves $D(E,t)$ in different energy bands $E$, one
derives count rate spectra $c_i(E)$ for the five components $i$. When we did
this independently for each of the 16 energy bins, we obtained quite noisy
spectra with a divergence towards non-plausible values in some channels. This is
due to significant degeneracy between the various components that we discuss in
\refsec{degeneracy}. A way to overcome this problem was to include a link in the
fitting between the values obtained in one energy bin and in some others. We did
this by adding an additional constraint to the $\chi^2$ minimization of the fit
so that the fitted parameter value $c_i$ would not be too far from an expected
value $c_i\umrm{exp}$ according to:
\begin{equation}
\label{eq:chi2}
\chi\dmrm{fit}^2 = \chi\dmrm{red}^2 + \frac{1}{\xi}\sqrt{
\sum_{i=1}^{5}{\left(\log{c_i}-\log{c_i\umrm{exp}}\right)^2} }\,,
\end{equation}
where $\chi\dmrm{red}^2\!\equiv\!\chi^2/\mbox{d.o.f.}$ is the $\chi^2$ divided
by the number of degrees of freedom (d.o.f.) and $\xi$ is a factor to be chosen
to get an appropriate balance between the quality of the fit and additional
constraints. We calculated the difference with respect to the expected value on
a logarithmic scale to avoid a dependence on the actual count rate values from
one component to the other.

The choice of the expected values $c_i\umrm{exp}$ in this constraint fit can of
course have strong implications on the results. We therefore took great care to
define them without including wrong assumptions and systematic effects. For the
parameter $c\dmrm{ins}$, the expected value was set to be the mean value
obtained over the 16 energy bins. This was motivated by our discussion in
\refsec{ins}, where we concluded that this factor can differ from unity, but is
expected to be rather constant from one energy bin to the other. For the four
other $c_i$ parameters, we defined the expected value based on the assumption
that the final, unfolded spectrum of each component is supposed to be smooth.
For each component spectrum, $c_i(E)$, the expected count rate in a given
channel, $c_i\umrm{exp}$, was set to be the linear interpolation between the
values in the two adjacent energy bins corrected for the effects of different
energy widths of the channels and of the detector response\footnote{To take into
account the effect of the ISGRI detector response, we used XSpec \citep{A96}
with the standard ISGRI ARF and redistribution matrix file (RMF) of OSA~7.0 to
derive the relative strength of the instrumental modulation of a powerlaw model
spectrum from one channel to the other. This modulation follows basically that
of the ARF, with some additional smoothing and distortion towards lower energies
due to the RMF, with only a slight dependence on the photon index taken to be
typically of $\Gamma\!=\!2$.}. For the first and last energy bins, the spectral
smoothness was similarly constrained, by setting the expected value to the
linear extrapolation of the two closest channels. \modif{We note that we did not
constrain the spectrum of the instrumental background level $a\dmrm{ins}$ and
its increasing or decreasing trend with time, $b\dmrm{ins}$, because both can
change rapidly from one energy bin to the other due to the presence of narrow
emission lines (see \refsec{data}).} For the five other parameters, the
introduced interdependence of the values obtained in adjacent energy bins
typically reduces the number of free parameters of the fit by a factor of 2.
This was taken into account when calculating the d.o.f. of the fit.

The actual fitting began with a set of input spectra and minimized the modified
$\chi^2$ of \refeq{chi2}, one channel after the other. We then reran this
process up to \modif{four} times until the overall $\chi^2$ -- computed on the
lightcurves in all energy bins -- did not significantly improve anymore. This
iterative spectral fitting was done independently for each of the four
Earth-observation datasets, \modif{and we combined the results to get the final
spectra.}

\modif{An issue in the fitting process is the choice of the parameter $\xi$
that} defines the strength of the additional constraint on the fit in
\refeq{chi2}. If $\xi$ is low the spectral smoothness constraint becomes
important and it then does not leave enough freedom to fit the actual data,
whereas if $\xi$ is too high, the fit might diverge in some energy bins towards
unrealistic values. \modif{We tested different values and chose $\xi\!=\!15$,
which leaves a lot of freedom to the fit, while limiting strong divergence to
only a few energy bins in one of the four EOs, namely EO\,3 in the 70--100\,keV
range (see \reffig{pltSpectra}).}

\modif{Another important issue is the choice of the input parameters, since we
experienced that they can have a significant influence on the final results.
This is due to degeneracy between some components that we discuss in
\refsec{degeneracy}. It is therefore safe to start with values corresponding
roughly to the expectations for the CXB and also for one of the two Earth
emission components.} We chose the analytical formula of \citet{GMP99} as the
basis for the input CXB spectrum. \modif{A first guess of the spectral values of
the other components was obtained} by doing a fit with the input CXB spectrum
fixed and imposing equal contributions for the two Earth emission components.
\modif{We obtained a global Earth emission that is much lower at the highest
energies than derived by \citet{CSR07}. As the CR-induced emission of the Earth
is the dominant component at these energies, this suggests that its
normalization has to be scaled down by a factor of $\sim$0.4 (see
\reffig{compEarIntegral}). The final set of unperturbated input spectra was
obtained by fitting the lightcurves again, but this time with fixed values for
both the CXB and the CR-induced Earth emission. For the latter, the spectrum was
defined by the analytical formula proposed by \citet[Eq.~(1)]{SCS07} with a
normalization $C$ of 13.2\,\keVcmssr, i.e. 0.4 times that derived by
\citet{CSR07}.}

To smear-out \modif{the dependence of the final results on these input spectra
and to estimate the uncertainties}, we made a series of fits with different
input parameter values. We did this by perturbating each channel value of the
initial spectra by a random deviation following a Gaussian distribution with a
$\sigma$ of 30\,\%. This was done independently for all five component spectra,
\modif{namely the CXB, the GRXE, the two Earth emission components and the
average level of the instrumental background.} We performed the whole fitting
process described above starting from 30 different sets of perturbated input
spectra. \modif{For the four EOs, this resulted in 120 spectral fits. Instead of
taking the average on the obtained values, we took the median in each energy bin
as a robust estimator of the mean. This has the advantage to be independent of
the use of the actual results or of their logarithm and is not influenced by
outstanding values. We estimated the 1-$\sigma$ (68\,\% CL) statistical
uncertainties on the median taking the two results at
$\pm34\,\%/\sqrt{4}\!=\!\pm17\,\%$ in rank order away from the median, where the
factor of four stands for} the four EOs being independent measurements.

\section{Results}
\label{sec:results}

\begin{figure}[t]
\includegraphics[width=\hsize]{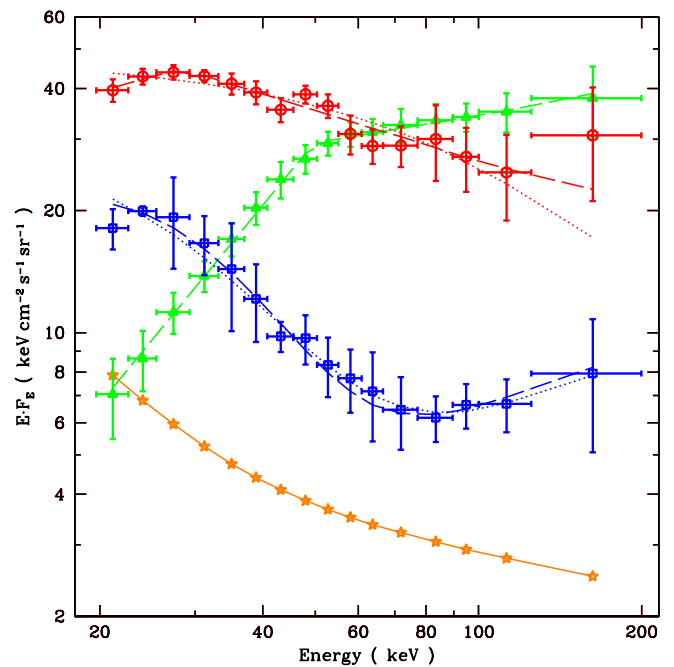}
\caption{\label{fig:plteeufs}
   Unfolded IBIS/ISGRI spectra of the sky background (red circles), the earth
   emission (green triangles) and the GRXE (blue squares) with their best fit 
   \modif{(dashed lines) and the more physical (dotted lines) spectral models
   (see \reftabs{results}{fitparam}). The contribution of the sum of the
   considered point sources averaged over the four EOs is also shown (orange
   stars).}
}
\end{figure}

\begin{table}[tb]
\caption{\label{tab:fitparam}
   Spectral fit parameters for the CXB, the GRXE and the Earth emission.
}
\begin{flushleft}
\begin{tabular}{@{}lcccccc@{}}
\hline
\hline
\rule[-0.5em]{0pt}{1.6em}
Spectra (model$^a$) & $\Gamma_1$$^b$ & $N_1$$^c$ & $E\dmrm{0}$$^d$ & $\Gamma_2$$^b$ & $N_2$$^c$ & $\chi\dmrm{red}^2$ \\
\hline
\rule{0pt}{1.2em}%
CXB (bknpow)		& 1.68		& 15.2	& 28.7	& 2.42		& --   & 0.51 \\
CXB (cutoff)$^e$	& 1.95$^f$	& 44.1	& 127	& --		& --   & 0.85 \\
GRXE (cutoff)		& 0.0$^f$ 	& 0.43	& 8.83	& 1.55$^f$	& 0.82 & 0.25 \\
GRXE (bremss)$^e$	& -- 		& 13.9	& 14.7	& 1.55$^f$ 	& 0.79 & 0.34 \\
Earth (bknpow)		& 0.37		& 0.050	& 49.4	& 1.78		& --   & 0.05 \\
\hline
\end{tabular}\\
\rule{0pt}{1.0em}%
\textbf{Notes.}
$^{(a)}$ Model is either a broken powerlaw (bknpow), a cutoff powerlaw
(cutoff) or bremsstrahlung (bremss), with or without an extra powerlaw.
$^{(b)}$ Photon index of low- (1) or high-energy (2) powerlaw.
$^{(c)}$ Normalization at 1\,keV in \phcmskeVsr.
$^{(d)}$ Characteristic energy in keV. Either the break energy (bknpow),
the cut-off energy (cutoff) or the $kT$ energy (bremss).
$^{(e)}$ the more physical model.
$^{(f)}$ fixed parameter value.
\end{flushleft}
\end{table}

The count rate spectra and uncertainties obtained with the iterative spectral
fitting process described in \refsec{spectral} are shown in \reffig{pltSpectra}.
\modif{We note that the large scatter from one EO to the other is clearly
dominating the uncertainties, which suggests that performing additional EOs in
the future will allow us to significantly improve the statistics of the
results.} We obtained overall average reduced $\chi^2$ values of
$\chi\dmrm{red}^2$\,=\,\modif{1.20, 1.11, 1.15 and 1.13} for EO\,1 to EO\,4,
respectively. These values only slightly above unity show that we get a fair
description of all the lightcurves, without overinterpreting the data by using
too many free parameters. As an example we show the match between the model
corresponding to the final results for EO\,1 and the observed lightcurves in
three representative channels in \reffig{pltFit}.

The count rate spectra $c_i(E)$ shown in \reffig{pltSpectra} correspond to the
values before entering the telescope, i.e. they are corrected for the
instrumental vignetting effects described in \refsec{instrumental}. We could
thus directly use them for spectral fitting with XSpec to get unfolded spectra
in physical units. We did the spectral fitting with the standard IBIS/ISGRI ARF
and RMF detector response files distributed with OSA~7.0. We did not consider
Galactic hydrogen absorption in the fit because even along the galactic plane
the hydrogen column density is small enough -- $N\dmrm{H}\!\approx\!2\times
10^{22}$\,cm$^{-2}$ at Galactic coordinates of ($l$,$b$)\,=\,(330\degr,0\degr)
-- to have only a negligible effect. The spectral fit parameters are given in
\reftab{fitparam} and the resulting unfolded spectra are shown in
\reffig{plteeufs} with the numerical values given in \reftab{results}.

\modif{The CXB spectrum is best fitted by a broken powerlaw model with a break
energy at $E\dmrm{b}\!=\!28.7$\,keV and a high-energy photon index of
$\Gamma_2\!=\!2.42\pm0.09$, giving a reduced chi-squared of
$\chi\dmrm{red}^2\!=\!0.51$ for 12 d.o.f. A more physical model for the CXB
emission -- considered as the superimposition of the emission of unresolved
Seyfert galaxies -- is to take a cut-off powerlaw model. The degeneracy between
spectral slope and cut-off energy was solved by fixing the photon index to the
value of $\Gamma\!=\!1.95$ derived by \citet{BSR09} on average for all Seyfert
galaxies detected by \emph{INTEGRAL}. We then obtained a good description of the
CXB spectrum $\chi\dmrm{red}^2/\mbox{d.o.f.}\!=\!0.85/14$ with a cut-off energy
of $E\!=\!127\pm20$\,keV, at slightly higher energy than $E\!=\!86$\,keV derived
for Seyfert~1 galaxies \citep{BSR09}.}

We found that the GRXE spectrum was best fitted by a cut-off powerlaw plus a
second powerlaw to account for the hard tail at energies above $\sim$\,80\,keV.
\modif{As the indices of the powerlaws are poorly constrained, we fixed their
values to $\Gamma_1\!=\!0.0$ for the cut-off powerlaw and to $\Gamma_2\!=\!1.55$
for the hard tail, as derived by \citet{BJR08}. This gives a very good
description of the data with a $\chi\dmrm{red}^2\!=\!0.25$ for 13 d.o.f.
According to \citet{RSG06}, the main population contributing to the low-energy
part of the GRXE are intermediate polar cataclysmic variables. The accretion
column onto the magnetic poles of such types of accreting white dwarfs is
emitting optically thin thermal emission. The best fit for this more physical
bremsstrahlung model is almost undistinguishable from the cut-off powerlaw (see
\reffig{plteeufs}) and gives a typical average temperature of
$kT\!=\!14.7\pm1.4$.}

\modif{The spectrum of the total Earth emission is best fitted by a broken
powerlaw with a break at $E\!=\!49.4\pm4.9$\,keV and a high-energy photon index
of $\Gamma_2\!=\!1.78\pm0.13$. This break energy is slightly higher than derived
by the recent analysis of the \emph{Swift}/BAT data by \citet{AGS08}, while the
obtained spectral slope is in remarkable agreement with their result of
$\Gamma_2\!=\!1.72\pm0.08$ (90\,\% CL errors). We found a different
normalization of the Earth emission spectrum however, which we discuss in
\refsec{earth}, where we also discuss the separate spectra obtained for the
albedo and the CR-induced emission.}

\modif{The spectrum of the sum of all point sources detected at more than
$2\,\sigma$ on average among the 4 EOs is added in \reffig{plteeufs} for
comparison. The impression that point sources contribute much less than the GRXE
is misleading. This is related to the arbitrary area we chose for the
normalization of the GRXE. If we had normalized it to the area actually covered
by the partially coded FoV of IBIS, we would have had a GRXE spectrum scaled
down by a factor of $\sim4$, depending a bit on the EO. This would then lead to
a higher contribution of the point sources compared to the GRXE in qualitative
agreement with the SPI results by \citet{BJR08}.}

\section{Discussion}
\label{sec:discussion}

The spectra in the $\sim$\,20--200\,keV range presented above will now be
compared to previously published \emph{INTEGRAL} results and to spectra obtained
by other satellites. In the subsections below, we discuss this separately for
the CXB spectrum, the GRXE and the Earth emission.

\begin{figure}[t]
\includegraphics[bb=16 144 600 510,clip,width=\hsize]{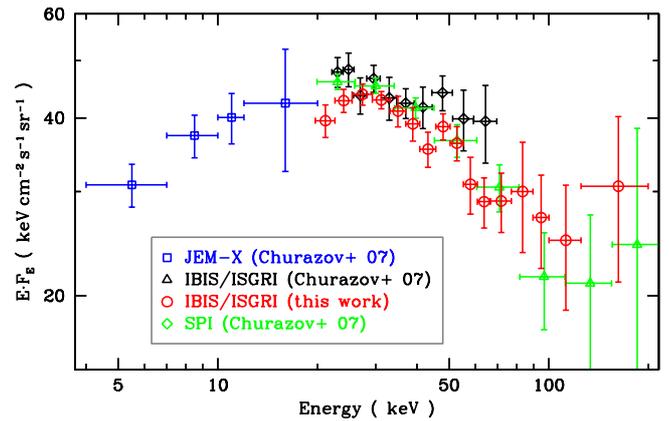}
\caption{\label{fig:compIntegral}
   Comparison of the IBIS/ISGRI CXB spectrum obtained here (red circles) with
   the previous \emph{INTEGRAL} results \modif{of IBIS/ISGRI (black diamonds),
   JEM-X (blue squares) and SPI (green triangles) published by \citet{CSR07}.}
}
\end{figure}

\subsection{Sky background spectrum}
\label{sec:sky}

It is interesting to compare the CXB spectrum obtained by the thorough analysis
of the IBIS/ISGRI detector lightcurves presented here with the \emph{INTEGRAL}
results previously published by \citet{CSR07}. The comparison is shown in
\reffig{compIntegral}. Our approach could significantly increase the useful
energy range of the IBIS/ISGRI data towards higher energies. \modif{The new
results fall slightly below the previous IBIS/ISGRI spectrum, while we get a
good agreement with the SPI results of \citet{CSR07}, except possibly for the
first energy bin.}

\begin{figure}[t]
\includegraphics[bb=16 144 600 510,clip,width=\hsize]{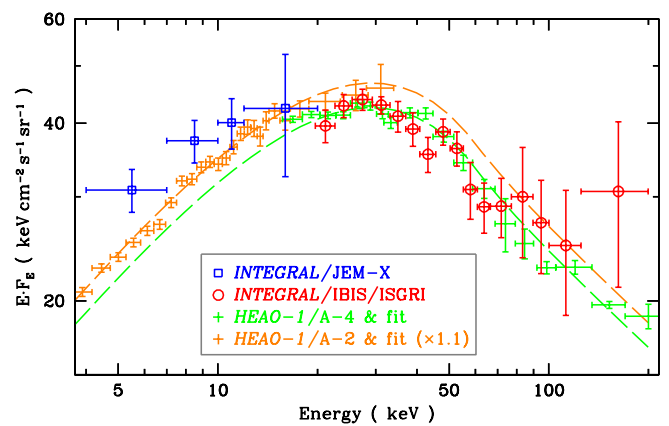}
\caption{\label{fig:compHEAO}
   \modif{Comparison of the CXB spectrum obtained by \emph{INTEGRAL} -- JEM-X
   measurements (blue squares) from \citet{CSR07} and our IBIS/ISGRI results
   (red circles) -- with the \emph{HEAO-1} spectra and analytical model by
   \citet{GMP99} (error bars and dashed line). The data of the A-4 instrument of
   \emph{HEAO-1} are shown in green with original normalization, while we show
   in orange the spectrum of the A-2 instrument and of the model, both increased
   by 10\,\% in intensity.}
}
\end{figure}

\begin{figure}[t]
\includegraphics[bb=16 144 600 510,clip,width=\hsize]{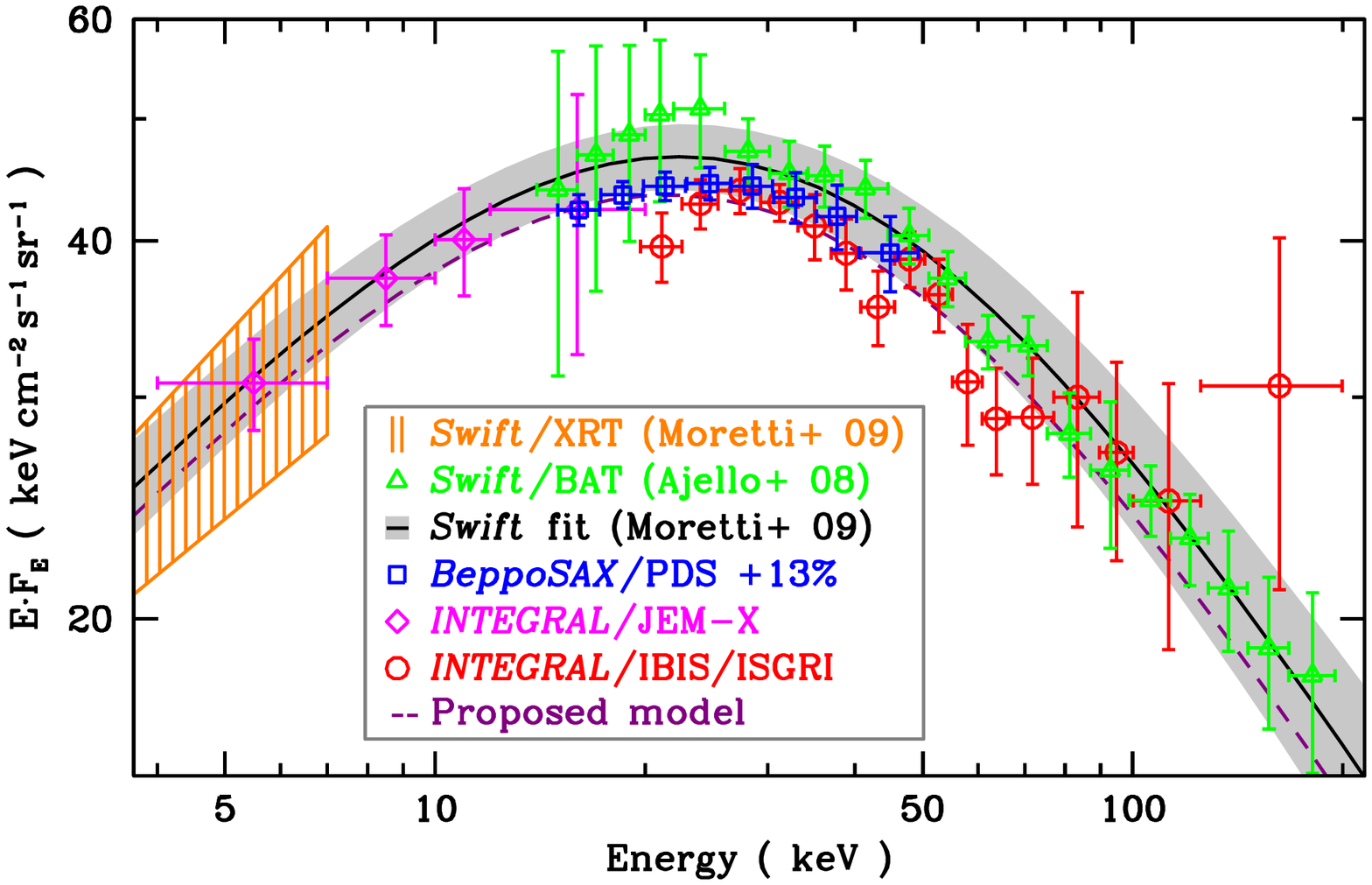}
\caption{\label{fig:compSwift}
   Comparison of the \emph{INTEGRAL} IBIS/ISGRI (red circles, this work) and
   JEM-X \citep[magenta diamonds,][]{CSR07} spectra with the other recent CXB
   measurements by \emph{Swift} and \emph{BeppoSAX}. The \emph{Swift}/XRT error
   box (orange shaded area) and the \emph{Swift}/BAT results (green triangles)
   are from \citet{MPC08} and \citet{AGS08}, respectively. \modif{The best-fit
   model of \citet{MPC08} for the combined \emph{Swift} dataset is shown with a
   black line and grey uncertainty area.} The original \emph{BeppoSAX}/PDS
   measurements of \citet{FOL07} were scaled by +13\,\% in intensity (blue
   squares) to correct for the difference in Crab normalization with respect to
   \emph{INTEGRAL}. \modif{The analytical model we propose in \refeq{cxb} is
   shown as a purple dashed line.}
}
\end{figure}

As shown in \reffig{compHEAO}, the slightly lower emission we obtain now with
\modif{IBIS/ISGRI} is consistent with the \emph{HEAO-1}
measurements and its analytical approximation by \citet{GMP99}. The flux scaling
of the \emph{HEAO-1} spectrum by $+10$\,\% as suggested by \citet{CSR07} is
actually not required anymore in the IBIS/ISGRI energy range. The discrepancy
appears only below 20\,keV for the \emph{INTEGRAL}/JEM-X data that indicate a
higher CXB intensity than the \emph{HEAO-1} measurements. It seems therefore
that a simple scaling in intensity of the historic \emph{HEAO-1} spectrum is not
able to consistently adjust the combined CXB measurements of \emph{INTEGRAL},
both below and above the turnover.

\modif{However, as illustrated in \reffig{compHEAO}, there is some freedom
within the uncertainties to scale up by $\sim10$\,\% the intensity of the
spectrum of the A-2 instrument of \emph{HEAO-1} without changing that of the A-4
experiment. This would better match the JEM-X measurements and other results by
recent X-ray instruments, which all suggest a higher intensity below 20\,keV
than obtained by \emph{HEAO-1}/A-2 \citep[e.g.][Fig.~15]{GCH07}. The net effect
would be a broadening of the CXB hump and a slight shift of its maximum towards
lower energies. The expected qualitative consequence for an AGN population
synthesis of the CXB would be a reduction of the contribution of the most highly
obscured AGN, in particular the Compton-thick ones \citep[e.g.][]{TUV09}.
Alternatively, it could also indicate a slightly stronger contribution from a
population of distant (redshifted) luminous AGN compared to the local population
\citep[e.g.][]{TU05}.}

\modif{Another consistency check of our results is to compare them to the recent
\emph{Swift} and \emph{BeppoSAX} measurements. This is illustrated in
\reffig{compSwift} where the \emph{Swift}/XRT  and the \emph{Swift}/BAT spectra
are from \citet{MPC08} and \citet{AGS08}, respectively. Our data are consistent
with the \emph{Swift}/BAT results and the combined XRT--BAT spectral model
proposed by \citet{MPC08}, although they tend to be at a significantly lower
intensity. Our data agree very well} with the \emph{BeppoSAX}/PDS data
\citep[Fig.~6 \emph{Bottom}]{FOL07} provided that they are scaled by a factor of
1.13 in intensity to account for the difference in the Crab normalization in the
20--50\,keV band between \emph{BeppoSAX}/PDS
\citep[$F\dmrm{Crab}\!=\!9.22\!\times\!10^{-9}$\ergcms]{FOL07} and
\emph{INTEGRAL} \citep[$F\dmrm{Crab}\!=\!10.4\!\times\!10^{-9}$\ergcms]{CSR07}.
We note that the latter estimation for \emph{INTEGRAL} is fully consistent with
the value obtained with OSA~7.0. The measured fluxes are
$F\dmrm{Crab}\!=\!10.30\!\times\!10^{-9}$\ergcms\ and
$10.46\!\times\!10^{-9}$\ergcms\ for the Crab observations of revolutions 365
and 422, respectively.
\modif{The obtained spectrum seems to be also very consistent in the peak region
with the recent CXB synthesis model by \citet{TUV09}. It thus gives additional
evidence for a small Compton-thick AGN fraction in the CXB spectrum, close to
9\,\% instead about 30--40\,\% postulated before \citep[e.g.][]{TU05}. Our data
cannot constrain a possible hardening of the CXB spectrum above 100\,keV, but
are consistent with an additional contribution of flat-spectrum radio quasars,
which have been found to dominate the CXB in the MeV range \citep{ACS09}.}

\modif{Based on the considerations above, we can tentatively suggest a slight
adaptation of the analytical description of the CXB proposed by
\citet[Eq.~(4)]{MPC08}, as:}
\begin{equation}
\label{eq:cxb}
E^2~\frac{dN_{\gamma}}{dE}=E^2~\frac{0.109~\phcmskeVsr}{(E/28\,\mbox{keV})^{1.40}+(E/28\,\mbox{keV})^{2.88}}\,,
\end{equation}
\modif{where the only difference -- but a correction of a typo in the units --
is a change of the break energy from 29\,keV to 28\,keV. The corresponding
spectral shape is at the lower limit of the uncertainty area of the \emph{Swift}
model as shown in \reffig{compSwift}.}

\subsection{Galactic ridge emission}
\label{sec:galactic}

\begin{figure}[t]
\includegraphics[bb=16 144 600 510,clip,width=\hsize]{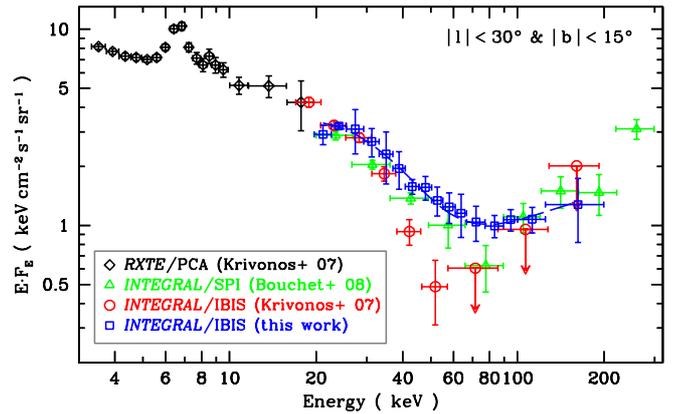}
\caption{\label{fig:compGal}
   Comparison of the obtained GRXE spectrum (blue squares)
   with recent other determinations, all renormalized to the central radian of
   the Milky Way defined by $|l|\!<\!30\degr$ and $|b|\!<\!15\degr$.
   The previous \emph{INTEGRAL}/IBIS data (red circles) and the \emph{RXTE}/PCA
   measurements (black diamonds) are from \citet[Fig.~14]{KRC07}. The
   \emph{INTEGRAL}/SPI spectrum (green triangles) is from \citet[Fig.~9]{BJR08}.
}
\end{figure}

It is not easy to compare results on the GRXE from one publication to the other,
because the emission is often defined in different regions of the Galaxy. As the
region covered by our observations is away from the Galactic bulge where most
determinations have been made, we have to rescale them to a more commonly used
area. We choose the central radian of the Milky Way defined in Galactic
longitude $l$ and latitude $b$ by $|l|\leq 30\degr$ and $|b|\leq 15\degr$ as the
reference area for a comparison of the various measurements. As we do have an
analytical model of the GRXE (see \refsec{spatial}), it is possible to determine
the scaling factor from any region in the Galaxy to the chosen area. The
resulting renormalized spectra are compared in \reffig{compGal}. Our
measurements had to be scaled by a factor of $0.16$ to correspond to the chosen
area. The \emph{INTEGRAL}/IBIS spectrum from \citet[Fig.~14]{KRC07}
corresponding to the IBIS FoV area centered on the Galactic bulge was
multiplied by a calculated factor of $2.77$. We used the same factor for the
\emph{RXTE}/PCA data that have been scaled by \citet{KRC07} to match the IBIS
measurement at 20\,keV. For the \emph{INTEGRAL}/SPI spectrum of
\citet[Fig.~9]{BJR08} that correspond already to the area chosen here, we just
had to convert the units.

In general, \reffig{compGal} shows a good agreement between the results of the
various instruments. This is quite remarkable for data that were not arbitrarily
renormalized, but were rescaled based on our very simple double-Lorentzian model
of the GRXE. This suggests that the model provides a fair description of the
overall emission of the inner Galaxy in the hard X-ray range. All three
independent \emph{INTEGRAL} measurements reveal a minimum at about 80\,keV. This
was only suggested by the 2-$\sigma$ upper limits of \citet{KRC07}, but is
confirmed now by our Earth occultation results and the latest SPI results of
\citet{BJR08}. \modif{Our IBIS/ISGRI results do, however, suggest that the
minimum is shallower than previously found.} The diffuse GRXE below 80\,keV is
thought to be due to a population of accreting white dwarfs too faint to be
resolved into discrete sources in the hard X-rays \citep{RSG06,KRC07}. It is
only at energies of $\sim$\,6--7\,keV that the diffuse emission could finally be
resolved using deep Chandra observations \citep{RSC09}. \modif{The
bremsstrahlung temperature of $kT\!=\!14.7\pm1.4$ that we derived in
\refsec{results} for the accretion column onto the pole of the white dwarfs
agrees well with the measurements of individual intermediate polar systems
detected by \emph{Swift}/BAT \citep{BGA09}. Based on Table~2 in the latter
publication, we note that this temperature would correspond to a typical white
dwarf mass of $M\dmrm{wd}\!\simeq\!0.60\pm0.05$\,M$_{\odot}$ according to the
model of \citet{SRR05}. This estimate is just slightly above the expected
average mass of white dwarfs in the Galaxy
($\overline{M\dmrm{wd}}\!\sim\!0.5$\,M$_{\odot}$) that was found to agree well
with the previous IBIS/ISGRI results on the GRXE \citep[and references
therein]{KRC07}.}

\begin{figure}[t]
\includegraphics[bb=16 144 600 510,clip,width=\hsize]{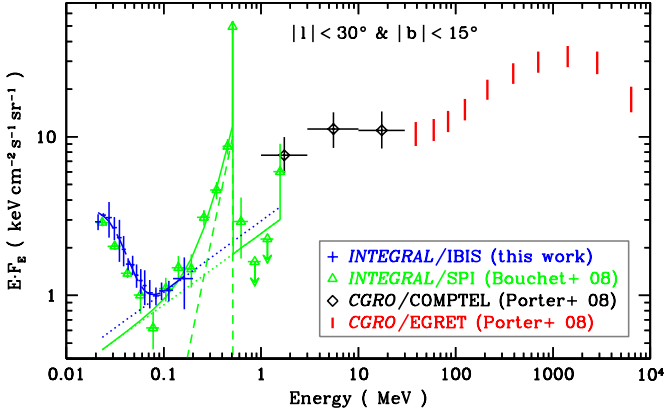}
\caption{\label{fig:compGal2}
   \modif{Comparison of the obtained GRXE spectrum (blue error bars and
   long-dashed line) with the higher-energy emission of the Galactic ridge in
   the region defined by $|l|\!<\!30\degr$ and $|b|\!<\!15\degr$ as observed by
   \emph{INTEGRAL}/SPI (green triangles) from \citet[Fig.~9]{BJR08} and by
   \emph{CGRO}/COMPTEL (black diamonds) and \emph{CGRO}/EGRET (red error bars).
   The \emph{CGRO} data are from \citet[Fig.~3]{PMS08}, which is based on the
   analysis by \citet{SBD99,SMR04}. The green solid line shows the model fitting
   the SPI observations above 100\,keV \citep{BJR08}. It is the sum of a
   powerlaw continuum (green dotted line), the positronium continuum
   (short-dashed line) and a narrow electron-positron annihilation line at
   511\,keV. A scaling by 20\,\% of the high-energy powerlaw is indicated by the
   blue dotted-line.}
}
\end{figure}

Above 80\,keV the GRXE spectrum is likely dominated by inverse-Compton emission
from the interstellar medium \citep{PMS08}. We derived \modif{an intensity at
the level of} the 2-$\sigma$ upper limits of \citet{KRC07}, \modif{in excellent
agreement with} the latest SPI observations \citep{BJR08}. \modif{The photon
index of the high-energy powerlaw derived by these authors ($\Gamma\!=\!1.55$)
also agrees very well with our results, although it is too poorly constrained by
our data alone to be fitted independently.}

\modif{The overall shape of the high-energy spectrum of the GRXE including the
observations of the Compton gamma-ray observatory (\emph{CGRO}) up to 10\,GeV is
shown in \reffig{compGal2}. As our data are scaled from a region at
$320\degr<l<340\degr$  lying outside the galactic bulge where the bulk of the
positronium annihilation is emitted, they should be almost unaffected by the
positronium continuum. This implies that the $\sim20$\,\%  higher normalization
of the high-energy powerlaw suggested by our data should be intrinsic. This
would agree well with the discussion of \citet{PMS08} concerning a possible
higher normalization of up to 40\,\% for this powerlaw.}

\subsection{Earth emission}
\label{sec:earth}

\begin{figure}[t]
\includegraphics[bb=16 144 600 510,clip,width=\hsize]{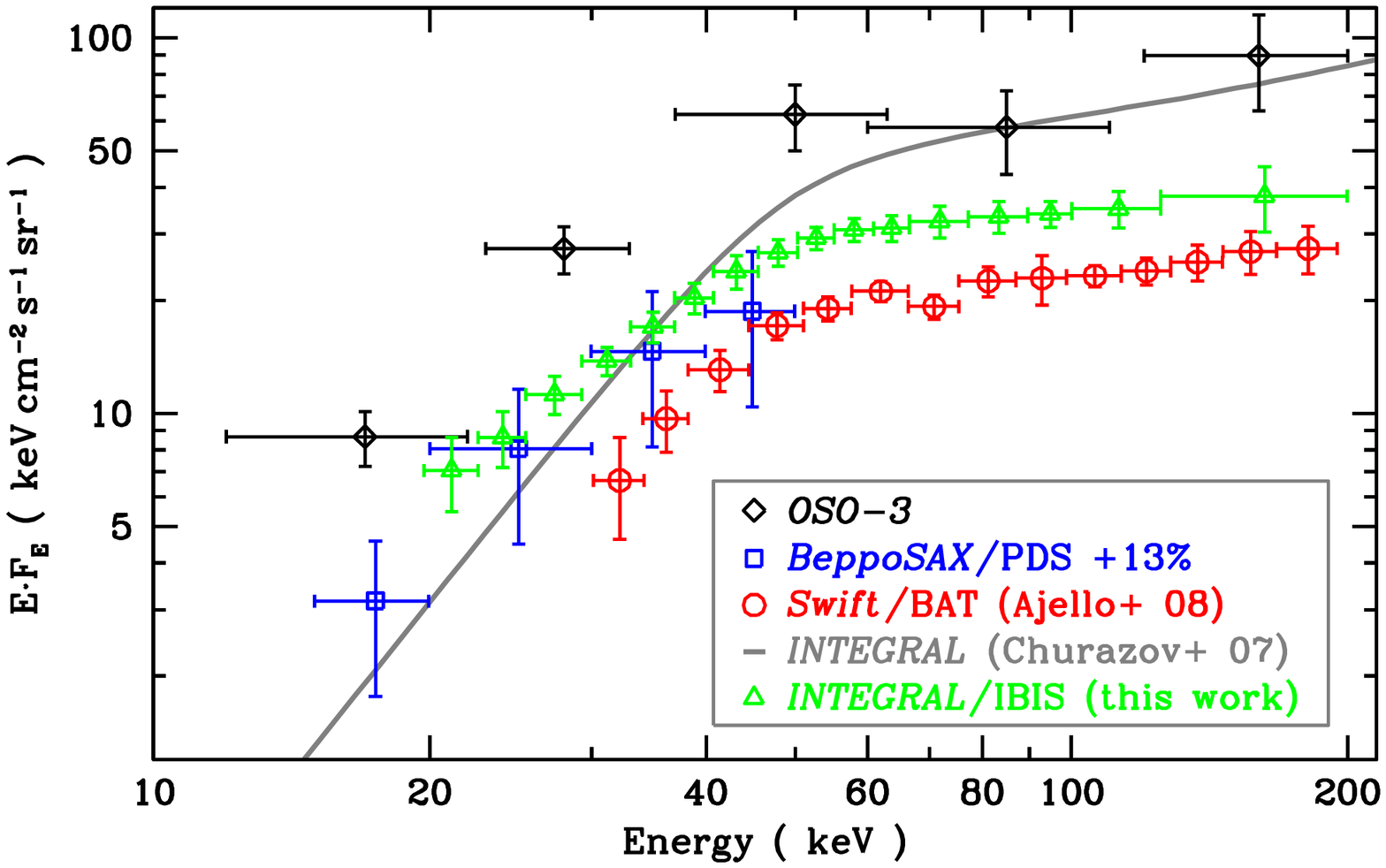}
\caption{\label{fig:compEarSwift}
   Comparison of the obtained Earth emission spectrum (green triangles) with
   previous determinations by various missions. The thin grey line is the
   \emph{INTEGRAL} spectrum of \citet{CSR07} as described in
   \reffig{compEarIntegral}. The obtained IBIS/ISGRI spectrum lies well between
   the \emph{OSO-3} (black diamonds) and the \emph{Swift}/BAT (red circles)
   measurements of \citet{SP74} and \citet{AGS08}, respectively. The values of
   the \emph{BeppoSAX}/PDS measurements of \citet{FOL07} were increased by
   13\,\% (blue squares) to be consistent with \reffig{compSwift}.
}
\end{figure}

\modif{Before discussing the relative contributions of the two Earth emission
components, we first compare, in \reffig{compEarSwift}, the obtained spectrum of
the Earth with other determinations. The Earth emission is found to be very
consistent with the spectra obtained previously, although there is a big scatter
among the various determinations. This is at least partially due to the
modulation of the Earth emission by the solar cycle and a dependence of the
observed flux on the spacecraft altitude and the geomagnetic latitude
\citep{SCS07}. For instance, the difference in normalization between the
\emph{Swift}/BAT spectrum and our determination can be related to
\emph{INTEGRAL} drifting towards an almost polar orbit, while \emph{Swift}
has a more equatorial orbit. The difference by roughly a factor of two depending
on the energy is consistent with the difference found by the polar-orbiting
satellite 1972-076B between the equatorial and the polar regions \citep{INR76}.
We did not include these spectra here for the sake of clarity, but they would be
compatible with} the other determinations provided that they are corrected for
unsubtracted CXB emission as shown by \citet[Figs.~15 and 16]{AGS08}.

\modif{A discrepancy we cannot ascribe to a different observation epoch or a
different viewpoint is the inconsistency of our results with the spectrum
derived from the same \emph{INTEGRAL} observations by \citet{CSR07}. We derive a
higher Earth emission at low energies and a lower intensity at high energies.
The discrepancy at the highest energies could somehow be due to the fact that
our results are based on IBIS data and their results on SPI, although both
instruments are well cross-calibrated. It is also possible that the difference
comes from our more detailed modeling of several instrumental effects described
in \refsec{instrumental}. At the lowest energies, the neglection of point source
emission by \citet{CSR07} is a likely cause of the discrepancy, since for a
given CXB we need more Earth emission to compensate for the occultation of point
sources.}

\begin{figure}[t]
\includegraphics[bb=16 144 600 510,clip,width=\hsize]{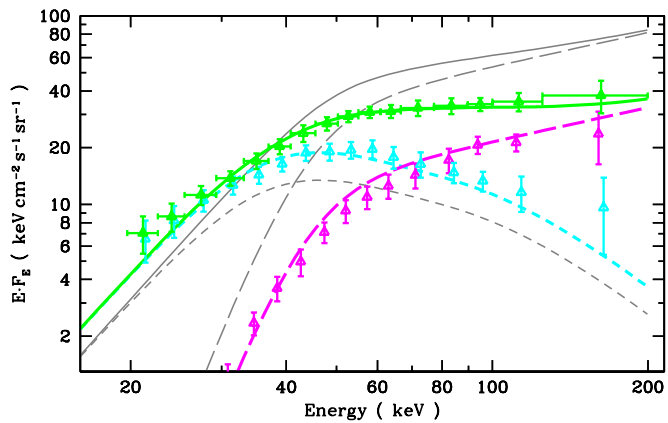}
\caption{\label{fig:compEarIntegral}
   \modif{Resulting spectrum of the total Earth emission (green triangles, solid
   line) with separated contributions from the Earth reflection of the CXB (cyan
   triangles, short-dashed line) and the CR-induced atmospheric emission
   (magenta triangles, long-dashed line). The thin grey curves are normalized as
   derived by \citet{CSR07}, whereas the thick colored lines are normalized to
   match our measurements.}
}
\end{figure}

\modif{To better characterize the difference between the two determinations of
the Earth spectrum, we show in \reffig{compEarIntegral} the decomposition of the
overall Earth emission in the two distinct components considered in both
studies. Those are the reflection of the CXB by the Earth -- the albedo -- and
the emission induced by CR interactions in the atmosphere. The spectra of both
components can be described by analytical functions fitted to the results of
Monte-Carlo simulations published by \citet{CSS08} for the albedo, and by
\citet{SCS07} for the atmospheric emission. In order to fit the overall spectrum
of the Earth with these two components, we need to increase the albedo component
by $\sim40$\,\% and decrease the atmospheric emission by $\sim60$\,\% compared
to the normalizations suggested by \citet{CSR07}. We thus get a normalization of
the atmospheric emission \citep[Eq.~(1)]{SCS07} of $13.2\,\keVcmssr$ at the
break energy of $E\!=\!44$\,keV, instead of $32.9\,\keVcmssr$ as derived by
\citet{CSR07}. This difference in normalization is much more important than the
few percent of likely overestimation mentioned by \citet{SCS07} related to the
inclusion of the Compton-scattering from particles that would have intersected
the surface of the Earth. Further speculations on the origin of this discrepancy
are beyond the scope of this paper.}

\modif{Concerning the albedo, we note that the increase by a factor $1.4$ we
derive here is not on the Earth reflection expected for the \emph{HEAO-1}
analytic approximation of \citet{GMP99}, but on this spectrum already scaled by a
factor of $1.1$ according to the results of \citet{CSR07}. As the CXB spectrum
obtained here agrees well with the original \emph{HEAO-1} spectrum, our
results suggest a reflection efficiency of the Earth atmosphere $\sim50$\,\%
higher than obtained by the Monte-Carlo simulations of \citet{CSS08}. According
to these authors, the shape of the input CXB spectrum has only a very limited
effect on the reflected spectrum, especially at energies below $\sim30$\,keV, so
there must be another reason for the important difference we observe. One
possibility is related to the delicate modeling of the composition of the
atmosphere. Could the presence of clouds have a significant effect by increasing
the amount of hydrogen atoms in the upper atmosphere, resulting in a more
``Sun-like'' albedo with more reflection at the lower energies
\citep[see][Fig.~6]{CSS08}? Another possibility would be the presence of
another Earth emission component emerging at the lowest energies, in particular
the potentially strong X-ray emission of aurorae \citep[e.g.][]{OSB01}. We note
that \citet{CSR07} found clear evidence for auroral emission in the
\emph{INTEGRAL}/JEM-X data during EO\,2 and EO\,3.}

\modif{Finally, it is fair to mention that the almost perfect agreement we show
in \reffig{compEarIntegral} between the data and the model for the two distinct
Earth emission components is mainly due to the choice of the input spectrum for
the atmospheric emission (see \refsec{spectral}). The strong degeneracy between
the two Earth emission components discussed in \refsec{degeneracy} does not
allow us to get such a good distinction of the two components when starting from
an arbitrary set of input parameters. We would just get a tendency for the
CR-induced emission to dominate at the higher energies and vice-versa at low
energy.}

\subsection{Degeneracy issues}
\label{sec:degeneracy}
Degeneracy is the reason why it is so difficult to determine the CXB spectrum
with Earth-occultation data. The basic problem is that a spatially uniform Earth
emission cannot be distinguished from the CXB occultation when the instrumental
background level is unknown. It is because of this that all previous studies
using the Earth as occultator had to assume a priori the spectrum of the CXB and
to a large extend also that of the Earth emission
\citep[see][]{CSR07,FOL07,AGS08}. Here, we tried to solve the degeneracy issue
by fixing the spatial distribution of the Earth emission components rather than
its spectral shape and by introducing a spectral smoothness constraint as
explained in \refsec{spectral}. However, because of the noise in the data and a
possible \modif{deviation} of the actual instrumental background variations
compared to those assumed based on the SPI/ACS lightcurve (see \refsec{ins}) it
is not possible to completely solve the degeneracy issues.

\modif{To overcome this problem, we had also to incorporate some a priori
assumptions on the spectral shape of the CXB and of one of the two Earth
emission components, chosen to be the atmospheric CR-induced emission because of
its strong drop at low energies that cannot be easily determined otherwise. As
explained in \refsec{spectral}, this is however only used to define the set of
input spectra that we then perturbate randomly before fitting the parameters to
the data. Despite this, we still keep a dependence on the input parameters in
the results. This is very obvious for the two Earth emission components that are
strongly degenerated (see \refsec{earth}).}

Another important degeneracy is between the instrumental background and the sum
of the sky background plus Earth emission. A higher instrumental background will
imply lower sky background and Earth emission, and vice-versa. Actually, the
data do primarily constrain the \emph{difference} between the Earth and the CXB
emissions. This difference is basically the height of a bump in the lightcurve
in case the Earth emission dominates or, alternatively, the depth of a trough,
when the sky background occultation dominates (see \reffig{pltFit}). \modif{To
test the effect of this degeneracy on our results, we fitted the data with the
same procedure as explained in \refsec{spectral}, but with an input spectrum for
the CXB increased by 10\,\%. This resulted in a CXB spectrum with a similar
shape, but a higher normalization by about the same factor. However, to
compensate this, the Earth albedo is then found to be higher than derived by
\citet{CSR07} by a factor of $\sim2$, instead of $\sim1.4$ (see \refsec{earth}).
Although this cannot be completely excluded, the discrepancy on the
normalization of the albedo is judged to be unrealistically high. We thus favor
the more conservative results obtained with the original normalization of the
CXB for the input parameters.}

The presence of the GRXE in the observed region of the sky can also add some
degeneracy, but its location to the side of the FoV was actually rather optimal
as its maximal occultation effect occurred earlier in the lightcurves than
for the CXB (see \reffig{pltLC}). \modif{We only noted a slight degeneracy
between the GRXE and the sum of the CXB and the polar-enhanced atmospheric
emission.}

\modif{We identified an instrumental effect that affects the results of
both the GRXE and of the Earth emission at energies above $\sim60$\,keV. By
ignoring the possibility of an increasing or decreasing trend of the
instrumental background in addition to the SPI/ACS modulation (see \refeq{fit}),
we obtained a high-energy drop of the Earth emission together with an
unrealistically steep rise of the GRXE spectrum. This behavior aims at
compensating a decreasing trend of the instrumental counts in the spectral
region of the emission lines of W, Pb and Bi (60--80\,keV, see \refsec{data})
and an increasing trend of the counts at even higher energies. This is likely
due to radioactive decay at the exit of the radiation belts and illustrates the
sensitivity of the method on a very accurate description of all instrumental
effects.}

\modif{Finally, we note that an underestimation of the contribution of the point
sources will tend to increase both the GRXE and the CXB intensity because point
sources are distributed all over the FoV with increased density in the galactic
plane.}

\section{Conclusion}
\label{sec:conclusion}

We presented the results of an original analysis of the four consecutive
Earth-occultation observations by the IBIS/ISGRI instrument aboard the
\emph{INTEGRAL} satellite. \modif{Our approach is complementary to the previous
study of these data by \citet{CSR07}, because instead of fixing the spectral
shape of the CXB and fitting its normalization, we attempt to derive the
complete spectral information from the observed detector lightcurves in
different energy bins. This requires a deep understanding of the instrumental
effects and a careful modeling  of the spatial distribution of the various
contributions from the Galaxy, the Earth and point sources.} Despite inherent
degeneracy issues that forced us to fit the data with an additional spectral
smoothness constraint and adequate input parameters, the approach used here
results in a coherent set of spectra for the CXB, the GRXE and the Earth
emission.

The obtained IBIS/ISGRI results for the CXB are consistent with the historic
\emph{HEAO-1} spectrum, without any scaling in intensity. \modif{The scaling by
$+10$\,\% in intensity proposed by \citet{CSR07} is not incompatible with the
actual dataset, but is disfavored as it implies a CXB reflection by the Earth
twice as strong than the one derived by the Monte-Carlo simulations of
\citet{CSS08}. The obtained spectrum also agrees well with recent \emph{Swift}
and \emph{BeppoSAX} determinations. We propose a slight adaptation of the CXB
model spectrum suggested by \citet{MPC08}, which is based on \emph{Swift} data
alone, that implies a reduced fraction of strongly absorbed AGN, compared to the
\emph{HEAO-1} spectrum of \citet{GMP99}.}

\modif{The spectrum of the Earth emission is very well described by the
contribution of two distinct components: the reflection of the CXB that is
dominant at lower energies, and the CR-induced atmospheric emission. The derived
normalizations for these two components is however found to be very different
from what was suggested by the study of \citet{CSR07}.}

With a total observation time of only about a day, these special types of
\emph{INTEGRAL} observations yield a spectrum of the GRXE with comparable
statistics as obtained by combining all available \emph{INTEGRAL}/SPI
observations. This allows us to observationally estimate the average mass of
white dwarfs in the Galaxy. Conducting similar observations in different regions
of the Galactic plane would be useful to characterize the \modif{longitudinal}
distribution of the GRXE.

\modif{However, it would be even more important to conduct Earth observations
away from the Galactic plane to lift any degeneracy related to the presence of
the GRXE and point sources. This would lead to a determination of the CXB with
improved statistics and less systematics and thus fully exploit INTEGRAL's
unique capability to observe the entire Earth from a high-altitude orbit.}

\begin{acknowledgements}
Based on observations with \emph{INTEGRAL}, an ESA mission with instruments and
science data centre funded by ESA member states (especially the PI countries:
Denmark, France, Germany, Italy Switzerland, Spain), Czech Republic and Poland,
and with the participation of Russia and the USA.
PL has been supported in part by the Polish MNiSW grants NN203065933 and
362/1/N-INTEGRAL/2008/09/0, and the Polish Astroparticle Network
621/E-78/BWSN-0068/2008.
\end{acknowledgements}

\bibliographystyle{aa}	
\bibliography{13072}	

\end{document}